\def\gtwid{\mathrel{\raise.3ex\hbox{$>$\kern-.75em\lower1ex\hbox{$\sim$}}}}
\def\ltwid{\mathrel{\raise.3ex\hbox{$<$\kern-.75em\lower1ex\hbox{$\sim$}}}}
\def\pmb#1{\setbox0=\hbox{#1}%
  \kern-.025em\copy0\kern-\wd0
  \kern.05em\copy0\kern-\wd0
  \kern-.025em\raise.0433em\box0 }
\def\brho{\pmb{$\rho$}}
\documentstyle[12pt,aasms4]{article}
\lefthead{GWINN ET AL.}
\righthead{VELA PULSAR SIZE}
\begin{document}
\input epsf

\title{SIZE OF THE VELA PULSAR'S EMISSION REGION AT 13~cm WAVELENGTH}
\author{
C.R. Gwinn$^1$, M.C. Britton$^{1,2}$,
J.E. Reynolds$^3$, D.L. Jauncey$^3$, E.A. King$^3$,
P. M. McCulloch$^4$, J.E.J. Lovell$^{4,5}$ 
C.S. Flanagan$^6$,
\& R.A. Preston$^7$
}

\par\noindent $^1$ Physics Department, University of California, 
Santa Barbara, California, 93106, USA
\par\noindent $^2$ School of Physics, 
University of Melbourne, Parkville 3052, Victoria, Australia
\par\noindent $^3$ Australia Telescope National Facility, 
Epping, New South Wales, 2121, Australia
\par\noindent $^4$ Physics Department, 
University of Tasmania, Hobart, 7001, Tasmania, Australia
\par\noindent $^5$ Institute of Space and Astronautical Science, 
3-1-1 Yoshinodai, Sagamihara, Kanagawa 229, Japan
\par\noindent $^6$ Hartebeesthoek Radio Astronomy Observatory,
Krugersdorp, Transvaal, South Africa
\par\noindent $^7$ Jet Propulsion Laboratory, California Institute
of Technology, Pasadena, California, 91109, USA
\vskip 1 truein
\begin{abstract}
We present measurements of the size of the Vela pulsar
in 3 gates across the pulse,
from observations of the distribution of
intensity.
We calculate the effects on this distribution
of noise in the observing system,
and measure and remove it using observations
of a strong continuum source.
We also calculate and remove the expected effects of
averaging in time and frequency.
We find that effects of 
variations in pulsar flux density and instrumental gain,
self-noise, and one-bit digitization are undetectably small.
Effects of normalization of the correlation 
are detectable, but do not affect the fitted size.
The size of the pulsar 
declines 
from $440\pm 90$~km (FWHM of best-fitting Gaussian
distribution) to less than 200~km across the pulse.
We discuss implications of this size for theories of pulsar emission.
\end{abstract}


\section{Introduction}

\subsection{Pulsar Radio Emission}\label{psr_emission}

Pulsars emit meter- to centimeter-wavelength radiation with 
brightness temperatures of $T>10^{23}$~K, greater than
any other astrophysical sources.  
The ultimate energy source for pulsar emission is
rotation of the neutron star. Combined with 
this rotation, the neutron star's magnetic field 
generates forces sufficient to
accelerate electrons from the stellar surface. 
These energetic electrons 
or the gamma-rays they emit can
pair-produce after a few
centimeters. 
The consequent cascade of electron-positron pairs
permeates the magnetic field of the pulsar out to the light cylinder.
The electrons and positrons follow the 
lines of the strong magnetic field closely.
Particles on the ``open'' field lines, which pass
through the light cylinder,
gradually carry away part or all of 
the rotational kinetic energy of the pulsar as an electron-positron wind,
leading to the observed spindown of the pulsar.
The ``polar cap'' is the region defined by the open field lines
at the surface of the neutron star,
from which the wind is drawn.
This basic picture has long been understood
(\cite{gol69}, \cite{stu71}, \cite{rud75}, \cite{aro83}).
Despite this understanding,
the physical mechanism for conversion of a small fraction
of the spindown power to a beam of radio emission,
which sweeps past the observer each rotation period
to produce the observed pulses, remains uncertain.
Observational tests of models for pulsar emission are difficult
because the emission is so compact.
It cannot be resolved with
instruments of Earthlike or smaller dimensions.

Melrose (1996)\markcite{mel96} and Asseo (1996)\markcite{ass96}
recently reviewed theories of pulsar emission.
In all models, collective behavior of
the electron-positron plasma results in coherent emission from
groups of particles.
Melrose divides theories into 4 categories,
according to location of the emission region:
(i) at a pair-production front at the polar cap;
(ii) in the electron-positron wind on open field lines above the polar cap;
(iii) in the electron-positron wind far above the cap, near the light cylinder;
and (iv) at, or even outside, the light cylinder.
Theories from all 4 categories appear viable.
Indeed, more than one emission mechanism may act, to
produce the wide range of 
observed pulse morphologies (\cite{ran83}).

Pulsars exhibit rich temporal structure,
including complicated variations of structure of individual pulses
about an average, and large pulse-to-pulse variations in flux density;
these offer clues to the emission process 
(\cite{han72}, \cite{lyn98}). 
Because the fundamental modes 
of lower-frequency electromagnetic waves
are linearly polarized, either across or along the pulsar's magnetic field
(\cite{bla76}, \cite{mel77}, \cite{aro86}, \cite{lyu98a}),
the strong linear polarizations of pulsars
yield the direction of the pulsar's
magnetic field at the locus of emission (\cite{rad69}). 
Polarization observations usually suggest that
emission arises near the magnetic pole of the neutron
star, which is offset from the rotation pole.

Pulsars show stable, sometimes complicated long-term
average profiles,
but with large pulse-to-pulse variations.
The most successful classification
of these profiles is that of
Rankin (\cite{ran83}, \cite{ran90}),
who divides emission into ``core'' and ``cone''
components,
corresponding respectively 
to elliptical pencil beams and hollow cones
directed from above the magnetic
pole of the neutron star.
A host of associated pulse properties support this classification 
(\cite{ran86}, \cite{lyn88}, \cite{rad90}).
The existence of these different components may 
indicate
the presence of more than one emission process
(\cite{lyn88}, \cite{wea97}).
A few core emitters show interpulses,
suggesting that their magnetic poles lie
near the rotation equator.
Comparison of pulse width with period
for these pulsars,
and extension of this relation
to all core emitters, closely
tracks the relation expected
if the size of the polar cap sets
the angular width of the beam for core emission (\cite{ran90}).
Rankin concludes that core emission arises very near the
surface of the neutron star.

Interstellar scattering offers the
possibility of measuring the sizes of pulsars and the structure of the
emission region, and its changes over the pulse,
using effective instruments of AU dimensions 
(\cite{bac75}, \cite{cor83}, 
\cite{wol87}, \cite{smi96}, \cite{gwi97}, \cite{IO98}).
In this paper, we describe measurement of the size of the Vela pulsar,
from the distribution of flux density in interstellar scintillation,
observed on a short interferometer baseline.

\subsection{Interstellar Scattering}\label{iss_intro}

Density fluctuations in the interstellar plasma scatter
radio waves from astrophysical sources.
These fluctuations act as a corrupt lens,
with typical aperture of about 1~AU.
Like such a lens,
scattering forms a diffraction pattern in the plane
of the observer.
For a spatially-incoherent source,
the intensity of the pattern is 
the convolution of the intensity of the pattern for a point
source, with an image of the source (Goodman 1968).
The scattering system has
resolution corresponding to
the diffraction-limited resolution of the ``scattering disk'',
the region from which the observer receives radiation.

The finite size of the scattering disk sets a minimum
spatial scale for the diffraction pattern in
the plane of the observer, at the diffraction limit.
This scale is $\lambda/\theta$,
where $\lambda$ is the observing wavelength and 
$\theta$ is the angular standard deviation of the scattering disk,
as seen by the observer.
If the plasma fluctuations are assumed to remain ``frozen''
in the medium, while the motions of pulsar, observer, and medium
carry the line of sight through it at speed $V_{\perp}$,
this spatial scale sets the timescale of scintillations,
$t_{ISS}=\lambda/\theta V_{\perp}$.
Differences in travel time from the different parts of the scattering
disk set the minimum frequency scale of the pattern,
at 
$\Delta\nu={{c}\over{2\pi}}
(\theta D)^{-2}
{\left[{{1}\over{D}}+{{1}\over{R}}\right]^{-1}}$,
where $D$ is the characteristic distance from observer to scatterer,
and $R$ is the characteristic distance from scatterer to source.
Observations with time resolution shorter than $t_{ISS}$
and with frequency resolution finer than $\Delta\nu$
are in the ``speckle limit'' of interstellar scattering.

For an extended, spatially-incoherent source,
overlap of different parts of the image of the source
in the observer plane, after the convolution,
reduces the depth of modulation of scintillation:
``stars twinkle, planets do not.''
The relation of Cohen et al. (1967) and Salpeter (1967)
quantifies this fact as a relation between the size of the source
and the depth of modulation of scintillations.

\subsection{The Vela Pulsar}

The Vela pulsar is particularly well-suited for studies of
pulsar emission because it is strong, heavily scattered,
and relatively nearby.
In a typical ``scintle'' of observing time $t_{ISS}$ and
bandwidth $\Delta\nu$,
a typical terrestrial radiotelescope can attain a signal-to-noise
ratio of a few, 
for observations of the Vela pulsar at decimeter wavelengths.
A short observation can thus sample many scintles.
Moreover, the scattering disk is large enough at these
wavelengths that the scale of the diffraction pattern
is on the order of the size of the Earth,
so that different radiotelescopes on Earth can sample different scintles.
The nominal linear resolution of the scattering disk,
acting as a lens, is about 1000~km at the pulsar.
For the Vela pulsar, the diameter of the light cylinder is 8500~km.
In contrast,
the size of the polar cap is 
tens of meters,
for a dipole magnetic field
and a $\sim 10$~km radius for the neutron star.
Thus, observations of the scintillation pattern can distinguish
among different theories of pulsar emission,
as discussed in \S \ref{psr_emission}.

Among the properties that place the Vela
pulsar's pulse into the core class are its single component, varying
little in width with frequency; its short period and rapid spindown
rate; its circular polarization; and its weak, irregular
pulse-to-pulse variations (\cite{ran83}, \cite{ran90}).  
However, Vela also shows characteristics
of cone emission, including strong, well-ordered linear polarization
and a double profile in X- and gamma rays (\cite{str96}).  
Indeed, Radhakrishnan et al. (1969)\markcite{rad69} first observed 
in the Vela pulsar
the S-shaped variation of
the direction of linear polarization characteristic
of cone pulsars.  
Manchester
(\cite{man95}) has proposed that Vela's pulse (and those of some other
young pulsars) could represent one side of broad cone, which would
help to explain the apparent inconsistency.  On the other hand, 
Romani \& Yadigaroglu (1995)\markcite{rom95} propose that the radio pulse is
indeed core emission, perhaps from very close to the polar cap, and
that the double-peaked 
X- and gamma-ray emission arises far out in the
magnetosphere, near the light cylinder.

Krishnamohan \& Downes (1983)\markcite{kri83} performed an extensive
study of 87040 pulses from the Vela pulsar, and found that the pulse
shape varied with peak intensity.  Notably, strong pulses tend to
arrive earlier.  They also found that the rate of change of angle of
linear polarization changes with pulse strength, and interpreted this
as activity from emission regions at different altitudes and magnetic
longitudes.  They inferred variations in the location of the emission
region by about 400~km in altitude, and by about about $4^{\circ}$ in
longitude.  Emission earlier in the pulse was found to arise further
from the star, and from a larger region.

\section{Distribution of Interferometric Correlation}\label{theory_section}

Interferometers present several advantages over single antennas
for observations of scintillating sources.
Because noise does not correlate between antennas,
no baseline response need be subtracted from observations.
Moreover, interferometers are nearly immune to interference
and emission from extended or unrelated nearby sources.
The statistics of interferometer response to signals and noise
is well studied
(see, for example, \cite{tms86}).
For a short interferometer baseline,
interferometers measure the intensity
of the source, commonly expressed as flux density.
Disadvantages of interferometers include the facts that
they require coordinated observations at multiple antennas,
and so often yield smaller effective apertures than the largest
available single antenna;
and they often
afford narrower bandwidths than some observing modes
at single antennas.
For the present work, careful treatment of statistics
is essential and interferometers are preferred.
We focus on observations of scintillating sources
on short interferometric baselines
in this paper.

\subsection{Distribution of Flux Density for a Small Source}

\begin{figure}[t]
% fig 1: 
\plotone{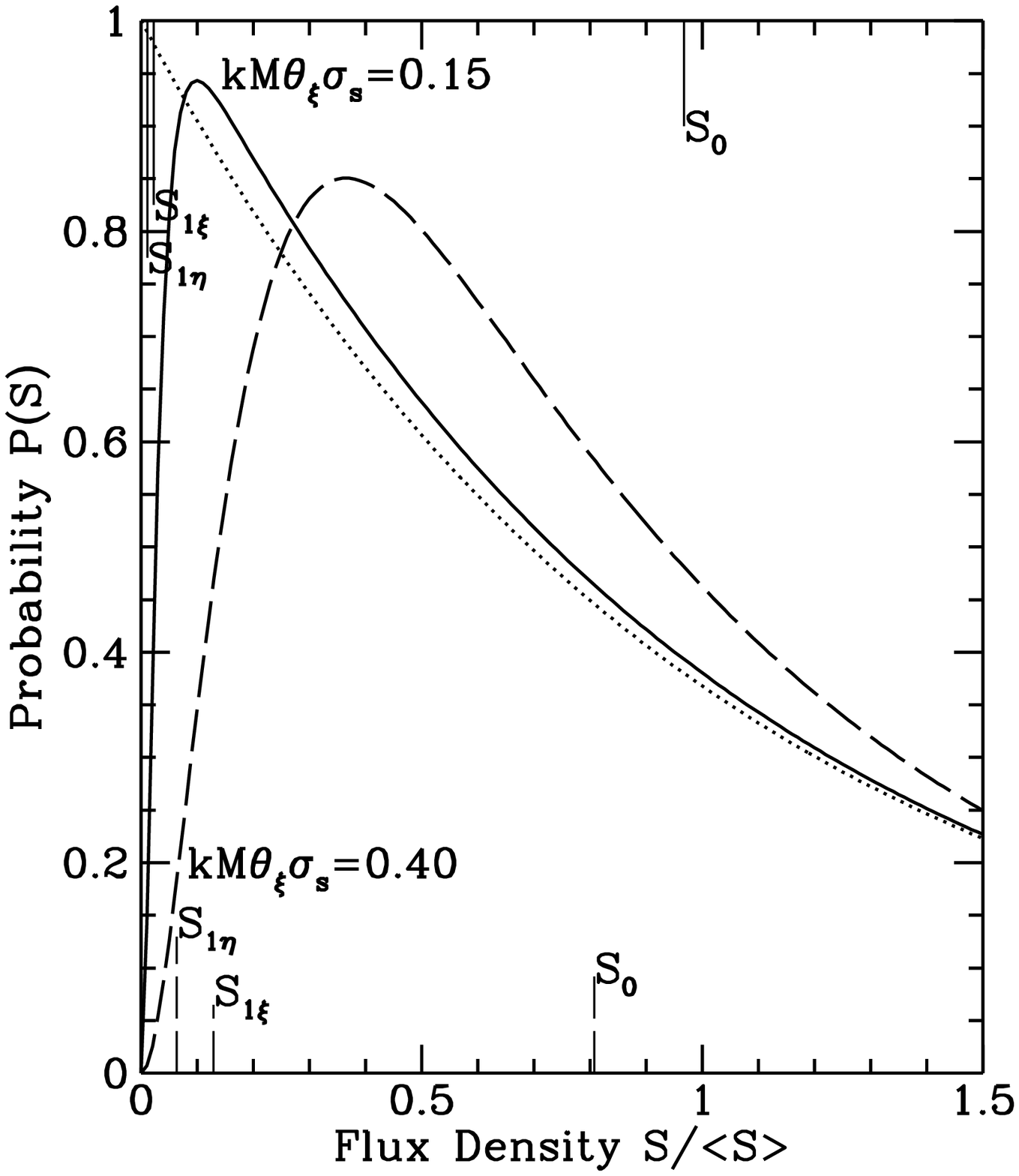}
\figcaption[]{
Probability distribution $P(S)$ for flux density $S$,
for a point source (dotted line),
and for sources with small size $\sigma_s$
($k M \theta_{\xi} \sigma_s = 0.15$: solid line)
and with larger size 
($k M \theta \sigma_s = 0.40$: long-dashed line).
Vertical lines 
near the top of the figure
indicate the scales $S_{1\xi}$, $S_{1\xi}$,
and $S_0$ for the smaller source;
those near the bottom indicate these scales for the larger source.
The mean flux density is the same for all 3 sources; $<S>=1$.
In these examples, 
the elongation of the scattering disk is $\theta_{\xi}/\theta_{\eta}=1.4$.
\label{3exp_fig}}
\end{figure}

For a point source in strong scintillation,
sampled in the speckle limit,
the distribution of flux density
follows an exponential distribution (\cite{sch68}, \cite{goo85}):
\begin{equation}
P(S)=1/S_0\, \exp\{-S/S_0\}.
\label{PSexpdist}
\end{equation}
This equation holds if the source is small,
compared with the resolution of the scattering disk
treated as a lens, so that
$kM\theta\sigma_s <<1$,
where 
$M=D/R$ is the magnification of that lens,
and $k=2\pi/\lambda$ where $\lambda$ is the observing wavelength.
The size of the source,
parametrized as the standard deviation of a
circular Gaussian distribution of intensity, is $\sigma_s$.
If $kM\theta\sigma$ is finite but small compared with 1,
then the distribution of flux density is the 
weighted sum of 3 exponentials (\cite{IO98}).
If, for example, the scattering disk is modeled as an elliptical Gaussian,
and the source is modeled as a circular Gaussian, then
the exponential scales are 
\begin{equation}
S_0,\quad
S_{1\xi}=S_0 (kM\theta_{\xi}\sigma_s)^2,
\quad{\rm and}\quad 
S_{1\eta}=S_0 (kM\theta_{\eta}\sigma_s)^2,
\label{Sscales_def}
\end{equation}
where $\theta_{\xi}$ and $\theta_{\eta}$ are the 
angular standard deviations
of the scattering disk along the major and minor axes,
and $S_0$ sets the scale of flux density.
We assume that
$(kM\theta\sigma_s)<<1$;
this is equivalent to the assertion that the scattering
disk, treated as a lens, does not resolve the source.
The probability distribution for 
flux density is then:
\begin{eqnarray}
P(S)
&=&{\textstyle
{{S_0}\over{(S_0-S_{1\xi})(S_0-S_{1\eta})}}
\exp{\left\{-S/S_0\right\}}
+{{S_{1\xi}}\over{(S_{1\xi}-S_0)(S_{1\xi}-S_{1\eta})}}
\exp{\left\{-S/S_{1\xi}\right\}}
}\label{PSell}\\
&&{\textstyle\quad 
+{{S_{1\eta}}\over{(S_{1\eta}-S_{1\xi})(S_{1\eta}-S_0)}}
\exp{\left\{-S/S_{1\eta}\right\}}.
}\nonumber
\label{S_3exp_eq}
\end{eqnarray}
At large flux denisty, the first term dominates
and the distribution remains exponential,
but with different scale $S_0$ and normalization
than for a point source.
At small flux density, the probability density 
falls to zero at zero flux density,
in a way that depends on source size $\sigma_s$.
The larger the source, the more sharply
the distribution of flux density 
is peaked near the average flux density.
Figure \ref{3exp_fig} shows sample distributions.
In this paper,
we use this expression,
with modifications for the
effects of noise, to find the size of the Vela pulsar
in 3 gates across the pulse.

Source size reduces the intensity variation due to scintillation.
The Cohen-Salpeter relation formalizes
this fact as a relation 
between angular size and 
modulation index, 
$m=\sqrt{\langle I^2\rangle-\langle I\rangle^2}/\langle I\rangle$.
Here the angular brackets $\langle...\rangle$ denote an ensemble average,
approximated in practice by a time average.
For a circular Gaussian source viewed through a 
elliptical scattering disk, in strong scattering (\cite{IO98}),
\begin{equation}
% m^2=(1+4 S_{1\xi}/S_0)^{-1/2}(1+4 S_{1\eta}/S_0)^{-1/2}.
m^2=(1+(2 k M \theta_{\eta}\sigma_s)^2)^{-1/2}
(1+(2 k M \theta_{\eta}\sigma_s)^2)^{-1/2}.
\label{modind_size}
\end{equation}
This expression is valid for arbitrarily large 
source size $\sigma_s$.
The modulation index calculated from
Eq.\ \ref{PSell} 
is consistent with this expression
in the limit $k M \theta_{\eta}\sigma_s<<1$.

\subsection{Effects of Averaging in Time and Frequency}\label{average_modind}

Because the scintillation pattern varies with time and frequency,
integration over either reduces the depth of modulation
and narrows the distribution of amplitude.
It thus affects the distribution in
a way similar to finite source size.
Averaging leaves the mean flux density $\langle S\rangle$ unchanged,
but reduces the mean square flux density $\langle S^2\rangle$,
and so reduces the modulation index.
We can calculate the effects of averaging on the modulation index,
and liken them to the effects of source size,
to assess effects of averaging on estimates
of source size.

For a source with pointlike or Gaussian intensity distribution,
seen through a Gaussian scattering disk,
the correlation flunction of flux density with frequency 
follows a Lorentzian distribution (see, for example, \cite{IO98}):
\begin{equation}
\langle S(\nu) S(\nu+d\nu)\rangle = 
\langle S(\nu)^2\rangle \left({{1}\over{1+(d\nu/\Delta\nu)^2}} + 1\right),
\label{Cnunupdnu}
\end{equation}
where $\nu$ and $\nu+d\nu$ are the observing frequencies,
and $\Delta\nu$ is the decorrelation bandwidth
of the scintillations.
For a small source with a Gaussian distribution
of intensity,
the correlation function has the same form,
with larger decorrelation bandwidth.
For non-Gaussian spectra of density fluctuations
in the scattering material,
the correlation function
shows the same behavior, with
a different functional form (\cite{cod86}).
After averaging in frequency
by convolving with a ``boxcar'' of bandwidth $B$,
the mean intensity is unchanged,
and the mean square intensity becomes
\begin{eqnarray}
\langle \bar S(\nu)^2\rangle &=&
\langle {{1}\over{B^2}}
\int_0^B dx\, 
\int_0^B dy\, 
S(\nu+x) S(\nu+y)\rangle \\
&=&{{2}\over{B}}
\int_0^B du\, (B-u)
\langle S(\nu) S(\nu+u)\rangle\label{freq_avg_s2}\nonumber \\
&=& \langle S(\nu)^2\rangle \left[
{{\tan^{-1}(B/\Delta\nu)}\over{B/\Delta\nu}}
+{{1}\over{2}}
-{{\ln(1+(B/\Delta\nu)^2)}\over{2 (B/\Delta\nu)^2}}
\right].\nonumber  \\
&\equiv& \langle S(\nu)^2\rangle f_f\nonumber
\end{eqnarray}
Here the bar on $\bar S$ denotes averaging.
This expression defines the convenient factor $f_f$.

A similar approach yields effects of averaging in time.
The spatial scintillation pattern
for a point source is
the Fourier transform of the 
screen phase (\cite{goo85}, \cite{cor89} \cite{IO98}).
Commonly, it is assumed that scintillation in time 
represents the motion of the scintillation pattern 
across the instrument, 
while the screen and pattern remain unchanged.
For an extended source,
the pattern is the convolution of 
the point-source response with an image of the source.
If both source and scattering disk follow Gaussian
distributions of intensity,
then the correlation function in time is Gaussian:
\begin{equation}
\langle S(t) S(t+dt)\rangle = 
\langle S(t)^2\rangle \exp\left\{
-\ln (2) (dt/t_{ISS})^2
\right\},
\end{equation}
where 
$t$ and $t+dt$ are the times of observation,
and $t_{ISS}$ is the timescale of interstellar scintillation.
If the observer averages over time interval $\tau$,
the mean square flux density after that averaging is
\begin{equation}
\langle {\bar S}^2\rangle = \langle S^2\rangle \left[
{\textstyle {{1}\over{\alpha}}}\sqrt{\pi}\,{\rm erf}(\alpha)+
{\textstyle {{1}\over{\alpha^2}}}e^{-\alpha^2}-
{\textstyle {{1}\over{\alpha^2}}}
\right]
\equiv \langle S^2\rangle f_t 
\label{time_avg_s2}
\end{equation}
where $\alpha=\sqrt{\ln 2}\;\tau/t_{ISS}$.

The effects of averaging can be expressed as
a reduction in modulation index or an increase in
inferred source size.
The post-averaging modulation index $\bar m$
is related to the pre-averaging value by 
\begin{equation}
\bar m^2=f_f f_t (m^2+1)-1.
\label{barm_def}
\end{equation}
Using Eq.\ \ref{modind_size}
we can infer a post-averaging
size $\bar\sigma_s$ from $\bar m$,
and so relate $\bar \sigma_s$ to 
the true size $\sigma_s$,
using knowledge of the averaging 
bandwidth and time.
This is the approach taken below.

\subsection{Noise and Self-Noise for Scintillating Sources}

The effects of noise can be important for studies of scintillating
pulsars because 
the scintles of lowest flux density are important in
determining the size of the source.
Similarly, noise affects the distribution of
intensity most strongly at the lowest intensities.
Thus, noise is important in determining source size.

\subsubsection{Noise and Self-Noise for Interferometric Correlation}

The interferometrist observes the signal from the source,
along with noise, at 2 stations $1$ and $2$ and measures
their correlation:
${{\bf C}_{12}}=C_x + i C_y
={{1}\over{N_q}}\sum_{p=1}^{N_q} ({\bf V}_{1p} {\bf V}_{2p}^*)$.
The correlation ${{\bf C}_{12}}$
is often called the interferometric visibility,
and is often expressed in units
of flux density, using the measured or estimated gains of the antennas.
For interferometric observations,
${{\bf C}_{12}}$ is complex (\cite{tms86}).
In this paper, we use boldface characters to denote complex quantities.
Here ${\bf V}_{1,2}$ are the electric fields in the observed
polarization, at the two antennas;
and $N_q$ is the number of independent samples, 
the product of integration time and bandwidth.
System noise 
may be in or out of phase with the signal,
and so contributes to both real and imaginary components.
Signals from astrophysical sources are intrinsically noiselike,
and their stochastic variations represent another
source of noise, known as self-noise.
Self-noise contributes only in phase with the signal.
Without loss of generality,
we take the phase of the signal to define the real axis.
Self-noise then affects only the real component.

For a strongly-scintillating point source,
the measured correlation ${{\bf C}_{12}}$ is drawn from the distribution:
\begin{equation}
P({\bf C}_{12})=\int du {{1}\over{2\pi \sigma^2}}
\exp\left\{ -{{1}\over{2}}{{(C_x-u)^2+C_y^2}\over{\sigma^2}}\right\}
\;\int dS\,f(u/S;N_q)
\;{{1}\over{S_0}}\exp\left\{-{{S}\over{S_0}}\right\},
\end{equation}
where $f(u/S; N_q)$ is the chi-square distribution
with $N_q$ degrees of freedom (see, for example, \cite{mey75}).
Again, $S_0$ is the mean flux density of the source,
averaged over many scintles.
We measure the correlation in units of flux density,
using the gains of the antennas $\Gamma_{1,2}$.
The system noise in units of flux density is
$\sigma^2=(T_{S1}/\Gamma_1)(T_{S2}/\Gamma_2)/N_q$
where 
$T_{S1,2}$ are the system temperatures at stations 1 and 2,
and $\Gamma_{1,2}$ are the gains at stations 1 and 2.
This equation assumes that the noise from the source
has Gaussian statistics.
In practice the radiation from 
all known astrophysical sources does appear to have Gaussian
statistics.
In particular,
radiation from pulsars follows Gaussian statistics closely,
with time-varying amplitude 
(\cite{ric75}).
For some pulsars, the amplitude 
varies quite dramatically over short times within the pulse
(\cite{han72}, \cite{lyn98}),
although the Vela pulsar seems to be relatively stable in this respect.

When the number of independent samples $N_q$ is large,
the chi-squared distribution is well approximated by
a Gaussian,
and the distribution takes the form:
\begin{equation}
P({\bf C}_{12})=\int_{0}^{+\infty} dS
{{1}\over{2\pi \sigma_x}} 
\exp\left\{-{{{1}\over{2}}}{{(C_x-S)^2}\over{\sigma_x^2}}\right\}
{{1}\over{2\pi \sigma_y}} 
\exp\left\{-{{{1}\over{2}}}{{C_y^2}\over{\sigma_y^2}}\right\}
\;{{1}\over{S_0}}\exp\left\{-{{S}\over{S_0}}\right\}.
\label{General_Cdist}
\end{equation}
The variances of the real and imaginary parts 
of the correlation are
\begin{equation}
\sigma_x^2={{{1}\over{N_q}}} {
(T_{S1}/\Gamma_1 +S)
(T_{S2}/\Gamma_2 +S)
}
\end{equation}
and 
\begin{equation}
\sigma_y^2={{{1}\over{N_q}}} {
(T_{S1}/\Gamma_1 )
(T_{S2}/\Gamma_2 )
}.
\end{equation}
The flux density 
of the source, $S$, 
appears in the denominator
of the exponent in Eq.\ \ref{General_Cdist},
through $\sigma_x$.
This reflects self-noise.

\begin{figure}[t]
% fig 2: from /home/cgwinn/vela/sim/dist_sim/self_noise/self_3.mac
\plotone{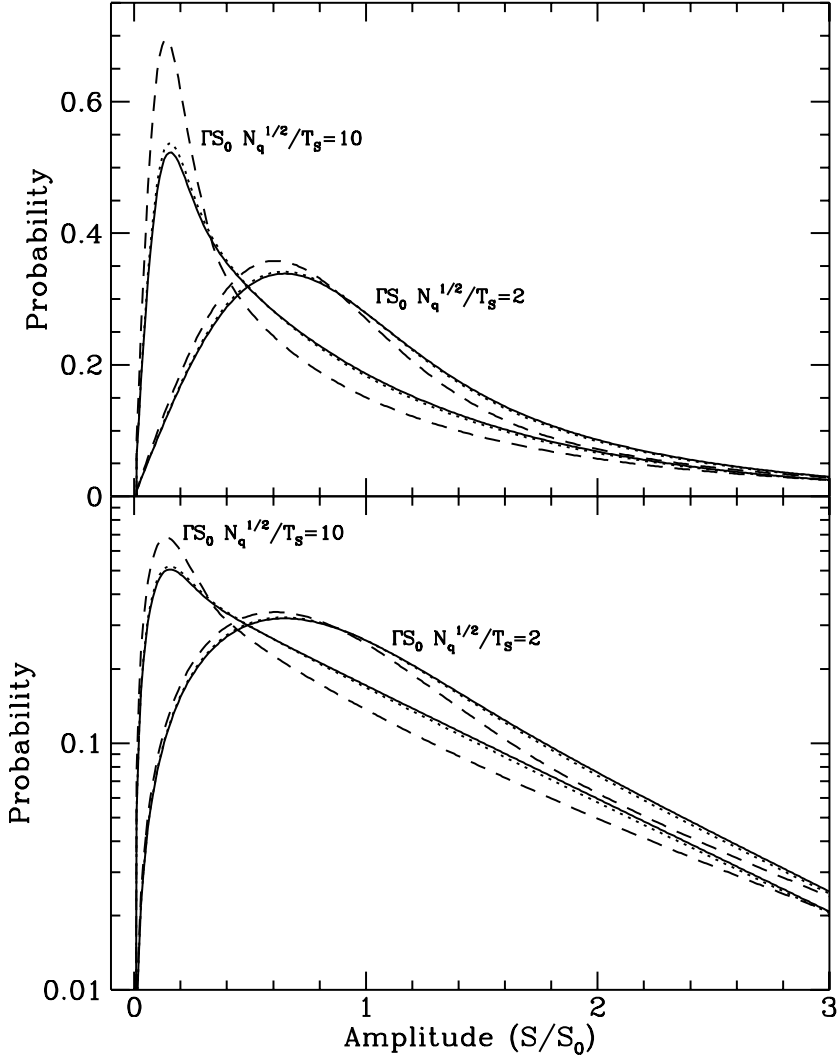}
\figcaption[]{
Upper panel: Distribution of amplitude of correlation, $|{\bf C}_{12}|$
for a scintillating point source, 
observed by 2 identical antennas, 
including effects of system noise and self-noise (see Eq.\ \ref{Cdist}).  
The mean flux density of the source is $S_0=1$.
The signal-to-noise ratio (excluding self-noise)
is $\Gamma S_0 N_q^{1/2}/T_s=10$ or $2$, 
as indicated for the 2 families of curves.
For each family,
the number of samples $N_q$ is $3000$ (solid line), $30$
(dotted line), and $10$ (dashed line).
The system temperature is $T_s$ and the antenna temperature is $\Gamma S_0$. 
Effects of
self-noise are greatest when the signal is large compared with
noise, and $N_q$ is small.
Lower panel: The same curves, plotted on a
logarithmic scale.  
\label{selfnoisedist_fig}}
\end{figure}

For 2 identical antennas, 
system temperature and gain are equal,
so that 
$T_{S1}=T_{S2}$ and $\Gamma_1=\Gamma_2$,
and
the correlation ${\bf C}_{12}=C_x+i C_y$
is drawn from the distribution:
\begin{eqnarray}
P({\bf C}_{12})&=&
{{N_q}\over{2\pi S_0 (T_S/\Gamma )}} 
\exp\left\{-{{(C_y)^2}\over{2(T_S/\Gamma )^2/N_q}}\right\}\label{Cdist}\\
&&\times
\int_{0}^{+\infty} dS 
{{1}\over{(T_S/\Gamma + S)}} 
\exp\left\{-{{(C_x-S)^2}\over{2(T_S/\Gamma + S)^2/N_q}}\right\}
\exp\left\{-S/S_0\right\}.
\nonumber
\end{eqnarray}
Unfortunately, this integral is not easily simplified.
Figure\ \ref{selfnoisedist_fig} shows plots of 
the distribution of the amplitude of the correlation,
$P(|{\bf C_{12}}|)$, 
for a scintillating point source,
for sample values of $\Gamma S_0/T_S$ and $N_q$.
When $N_q$ is large, self-noise
is negligible and the ratio $S_0\Gamma (N_q)^{1/2}/T_S$
is the average signal-to-noise ratio.
When $N_q$ is small,
self-noise becomes important.
As the figure shows, self-noise 
increases the number of points at low amplitude,
and flattens the distribution at high amplitude.

\subsubsection{Distribution for a Point Source with Noise}

If the number of independent samples $N_q$ is large,
then self-noise can be ignored.
The distribution of amplitude, $C_{12}=|{\bf C}_{12}|$, 
then takes the often-useful form:
\begin{eqnarray}
P_{S}(C_{12},S_0,\sigma)&=&P(u)\times {{du}\over{dC_{12}}}\label{erfC} \\
&=&{{1}\over{\sqrt{\pi}}}{{1}\over{S_0}}
\exp({\textstyle{{1}\over{4}}}\beta^2)\,u \nonumber \\
&&\quad \times \int_0^{\pi}\, d\phi
\exp\{-u^2\sin^2\phi - \beta u\cos\phi\}
\left(1+{\rm erf}(u\cos\phi - {\textstyle{{1}\over{2}}}\beta)\right)
\times {{1}\over{\sqrt{2}\sigma}},
\nonumber
\end{eqnarray}
where ${\rm erf}(~)$ is the error function (\cite{mey75}). 
The noise, in flux density units, is
$\sigma\equiv\sigma_x=\sigma_y=(T_{S}/\Gamma )$.
We have adopted the scaled parameters
\begin{equation}
u=C_{12}/\sqrt{2}\sigma\quad{\rm and}\quad \beta=\sqrt{2}\sigma/S_0.
\end{equation}
If the antennas are not identical,
the distribution has the same form in this limit,
but the definitions of the scaled parameters are slightly different.
Because self-noise can often be ignored,
as Figure\ \ref{selfnoisedist_fig} suggests,
this expression is often appropriate.
For scintles much stronger than noise
($C_{12}>>\sigma$),
and average flux density not much smaller than noise
(so that $S_0 >>\sigma^2/C_{12}$),
this distribution approaches
the purely-exponential form
\begin{equation}
P(C_{12})\approx\left\{\exp\left({{\sigma^2}\over{2S_0^2}}\right)\right\}
{{1}\over{S_0}}e^{-C_{12}/S_0}.
\label{erfC_approx}
\end{equation}
The constant of the exponential is the same as that for a
noise-free scintillating point source,
but the normalization is different.
In effect, noise shifts the distribution toward greater
amplitude.

\subsubsection{Distribution for a Small Source with Noise}

When the source has small but finite size,
and self-noise can be ignored,
the distribution of $C_{12}$ takes the form
suggested by the combination of Eqs.\ \ref{PSell} and\ \ref{erfC}:
\begin{eqnarray}
P(C_{12})
&=&{\textstyle 
{{S_0}\over{(S_0-S_{1\xi})(S_0-S_{1\eta})}}
P_S(C_{12},S_0,\sigma)
+{{S_{1\xi}}\over{(S_{1\xi}-S_0)(S_{1\xi}-S_{1\eta})}}
P_S(C_{12},S_{1\xi},\sigma)
}\label{3Cdist} \\
&&{\textstyle\quad 
+{{S_{1\eta}}\over{(S_{1\eta}-S_{1\xi})(S_{1\eta}-S_0)}}
P_S(C_{12},S_{1\eta},\sigma),
}\nonumber
\end{eqnarray}
Here Eq.\ \ref{Sscales_def} defines $S_0$, $S_{1\xi}$, and $S_{1\eta}$,
and Eq.\ \ref{erfC} defines $P_S(C_{12},S_n,\sigma$).
This relatively simple relationship holds 
even though the effects of noise on the amplitude distribution
cannot be described as a convolution.
For scintles with amplitude much larger than nosie,
and average flux density not much less than noise,
($C_{12}>>\sigma$ and $S_0 >>\sigma^2/C_{12}$,
the limits of Eq.~\ref{erfC_approx}),
only the first of the 3 terms is important, and 
this distribution also approaches the purely-exponential form:
\begin{equation}
P(C_{12})\approx\left\{\exp\left({{\sigma^2}\over{2S_0^2}}\right)\right\}
\left\{
(1-(kM\theta_{\xi}\sigma_S)^2)
(1-(kM\theta_{\eta}\sigma_S)^2)
\right\}^{-1}
{{1}\over{S_0}}e^{-C_{12}/S_0}.
\label{3Cdist_approx}
\end{equation}
In this expression, 
effects of source size appear only through the normalization.
This fact demonstrates that source size affects the
form of the distribution only at small amplitude.

\subsection{Effects of Variations in Flux Density and Instrumental Sensitivity}\label{variations_gain_flux_sec}

\begin{figure}[t]
% fig 3: 
\plotone{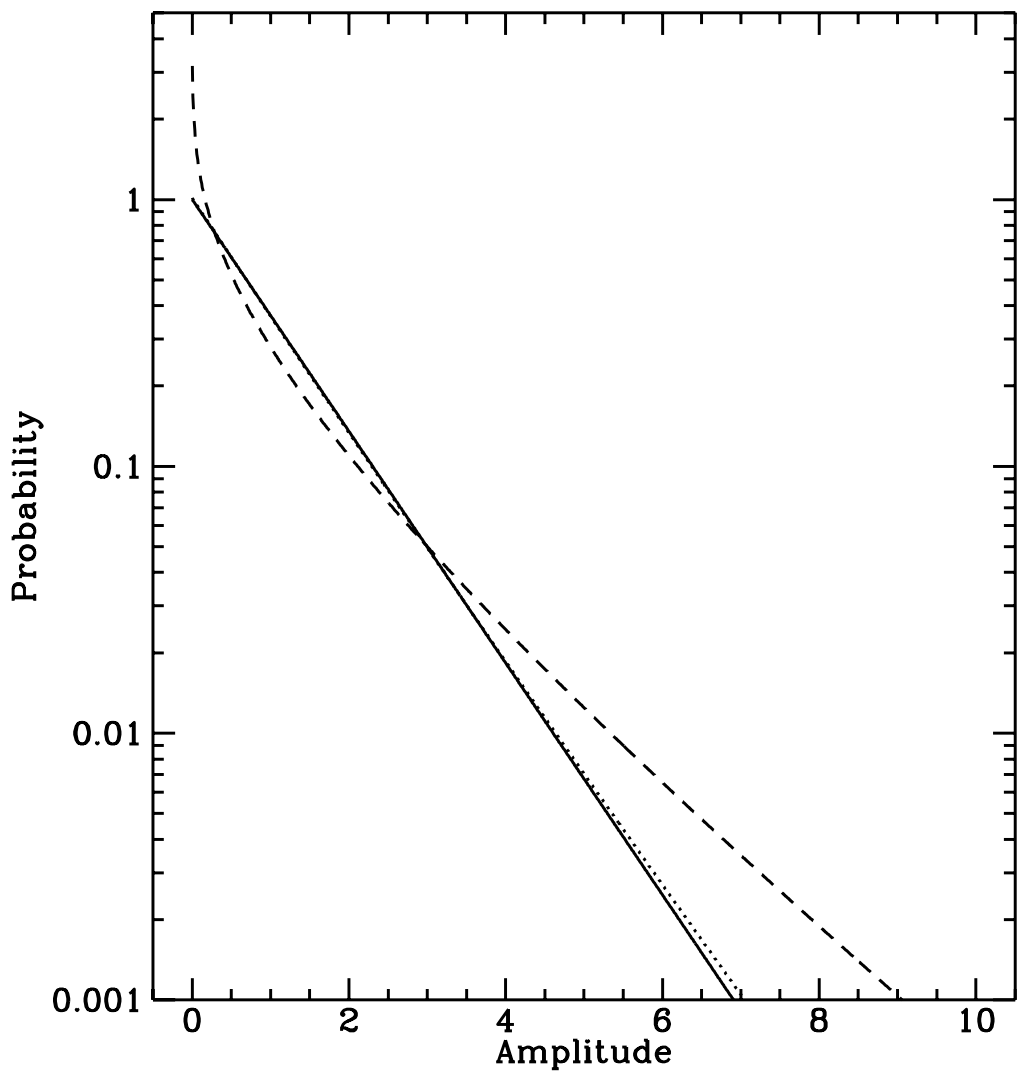}
\figcaption[]{
Expected distributions of amplitude of
correlation for a scintillating point
source, including effects of variations in instrumental gain (or, equivalently,
intrinsic variations in flux density).  
The solid curve shows the exponential distribution
expected for a point source in strong scintillation,
with mean and exponential scale equal to the 
mean amplitude of the source,
1 in this example.
Variations in gain change this distribution by
superposing distributions with different
exponential scales.
The model distributions of gain are flat,
extending from 0.8 to 1.2 times the mean (dotted curve)
and from 0 to 2 times the mean (dashed curve).
Distributions with significant contributions at zero or very low
gain differ the most from the 
case of constant gain.
\label{superposed_dist_fig}}
\end{figure}

If the gain of the observer's instrument varies with frequency or time,
then the distribution of observed flux density
will be the superposition of several distributions of the 
form given by Eq.\ \ref{PSexpdist} (or Eq. \ref{PSell},
if the source is extended).
For example, consider a scintillating point source
observed with a distribution of gains
$f(\Gamma)$, where $\Gamma$ is the gain.
We take 
$S$ to be the true flux density of the source,
and $T_A=\Gamma S$ to be the antenna temperature,
the observed quantity.
The distribution of $T_A$ is 
then the convolution of the distribution of $S$ with that for
$f$:
\begin{eqnarray}
P(T_A)&=&\int_0^\infty d\Gamma f(\Gamma) \int_0^\infty dS P(S) 
\delta(\Gamma S-T_A)\\
&=&\int_0^\infty d\Gamma f(\Gamma) {{1}\over{S_0\Gamma}} \exp(-T_A/\Gamma S_0).
\nonumber
\label{superposed_dists}
\end{eqnarray}
Similarly,
if the flux density of the source is not constant,
then the distribution of measurements 
includes contributions from exponential distributions 
with different scales.
The overall distribution of flux density will be the same
(Eq.\ \ref{superposed_dists}),
although the effective gain variations
will be stochastic rather than deterministic.
In either case, the distribution function will 
be concave upward on a semilog scale,
because the distributions with highest average flux density
fall off most slowly.
Figure\ \ref{superposed_dist_fig} shows examples,
for flat distributions of gain about the mean.
The effect is significant when the distribution gains 
includes very small or zero values.

\subsection{Normalized Correlation and the Van Vleck Relation for Scintillating Sources}\label{normcorr_vanvleck_sec}

Practical interferometers measure approximations
to the visibility ${\bf C}_{12}$.
In particular, interferometers almost always measure
the normalized correlation (\cite{tms86}, pp. 214, 248):
\begin{equation}
{\brho}_0=\eta {{\bf C}_{12}
\over{\sqrt{(T_{S1}/\Gamma_1+S)(T_{S2}/\Gamma_2+S)}}}.
\label{rho_0_TA_TS}
\end{equation}
Here $\eta$ is a correlator-dependent constant,
and $S$ is the flux density of the source.
Note that the denominator in Eq.\ \ref{rho_0_TA_TS}
does not include the number of samples $N_q$.
The denominator is simply a normalization factor; it does
not arise from, or characterize, the noise in the measurement.

Most signals are digitized before correlation.
For 1-bit sampling, as for 
the Haystack correlator used for the observations described here,
the measured correlation coefficient
$\rho_2$ is related to $\rho_0$ by the Van Vleck relation:
\begin{equation}
\rho_2={2\over{\pi}}\sin^{-1}\rho_0.
\label{van_vleck_relation}
\end{equation}
Often correlators 
tabulate the correlation coefficient 
scaled by the linearized Van Vleck correction
${\textstyle{{\pi}\over{2}}}\rho_2$.
Because normalization and digitization become
nonlinear at large amplitude,
and noise and source structure are important at small amplitude,
their effects are relatively easily separable.

Because normalization and digitization are important at large
amplitude, 
where the distribution of amplitude is exponential even
in the presence of source structure and noise,
we investigate effects of normalization and digitization on such
a distribution: $P(S)=(1/S_0)\exp\{-S/S_0\}$.
For 2 identical antennas observing an unresolved source,
${\brho}_0={{\bf C}_{12}/{(T_S/\Gamma+S)}}$,
and visbility is equal to the flux density, $C_{12}=S$.
Then the distribution of $\rho_0$ will be
\begin{equation}
P(\rho_0)=
{{T_S}\over{\Gamma S_0 \eta}}
{{1}\over{(1-\rho_0/\eta)^2}}
\exp\left\{-{{T_S}\over{\Gamma S_0}}{{\rho_0/\eta}\over{1-\rho_0/\eta}}\right\}.
\label{rhodist_eq}
\end{equation}
Note that the characteristic scale of the exponential
is the signal-to-noise ratio in a single sample, $\Gamma S_0/T_S$.
It is independent of integration bandwidth or time.
At small $\rho_0$,
$P(\rho_0)$ can be approximated by an exponential distribution:
\begin{equation}
P(\rho_0)\approx {{T_S/\Gamma }\over{S_0 \eta}}
\exp\left\{-\left({{T_S/\Gamma }\over{S_0}}-2\right)\rho_0 \right\}.
\label{rhodist_eq_approx}
\end{equation}
The additive constant ``2'' in the exponential arises from the factor of
$(1-\rho_0/\eta)^{-2}$ before the exponential in eq.\ \ref{rhodist_eq}.

\begin{figure}[t]
% fig 4: 
\plotone{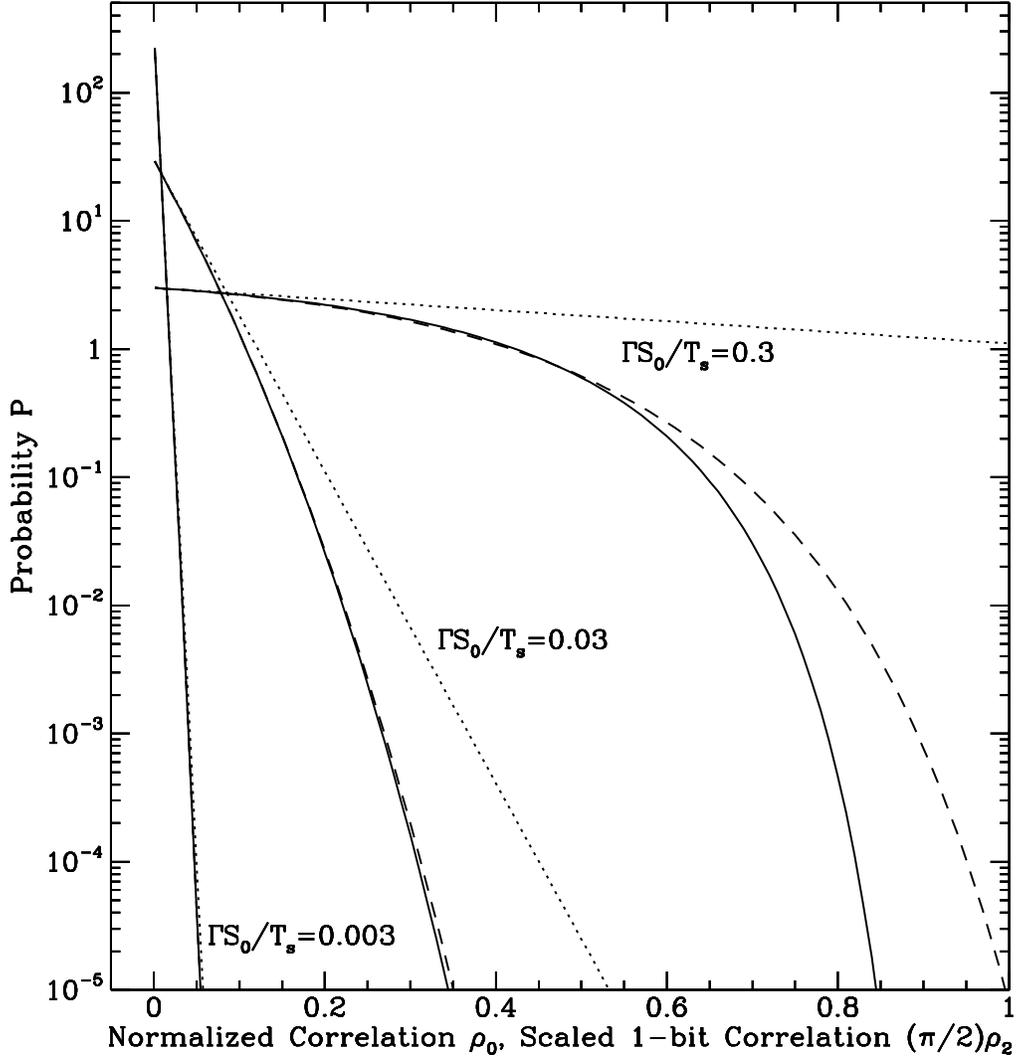}
% from rho.mac
\figcaption[]{
Probability distribution
of normalized correlation $\rho_0$,
and scaled 2-bit correlation ${\textstyle{{\pi}\over{2}}}\rho_2$,
for a strongly-scintillating source
with exponential distribution of flux density.
The average flux density, and scale of the exponential distribution,
is $S_0$.
Antenna temperature is $\Gamma S_0$, and system temperature is $T_S$,
so $\Gamma S_0/T_S$ is signal-to-noise ratio
for a single sample.
For each of 3 values of $\Gamma S_0/T_s$,
the probability distribution is shown for 
the normalized correlation $\rho_0$ (solid curve: Eq.\ \ref{rhodist_eq});
the scaled 1-bit correlation ${\textstyle{{\pi}\over{2}}}\rho_2$ 
(dashed curve: Eq.\ \ref{pio2rho2dist_eq}, with $\eta=1$);
and the limiting exponential form for small correlation
(dotted curve: Eq.\ \ref{rhodist_eq_approx}).
The curves differ only when the source is strong
relative to system noise.
All forms assume that $N_q$ is large
so that self-noise can be ignored.
\label{rhodist_fig}}
\end{figure}

We can also include the effects of 1-bit sampling,
by including the Van Vleck relation in calculating
this distribution.
For the scaled correlation coefficient,
\begin{equation}
P({\textstyle{{\pi}\over{2}}}\rho_2)={{T_S/\Gamma }\over{S_0}}
{{\cos({\textstyle{{\pi}\over{2}}}\rho_2)/\eta}
\over{(1-\sin({\textstyle{{\pi}\over{2}}}\rho_2)/\eta)^2}}
\exp\left\{-{{T_S/\Gamma }\over{S_0}}
{{\sin({\textstyle{{\pi}\over{2}}}\rho_2)/\eta}
\over{1-\sin({\textstyle{{\pi}\over{2}}}\rho_2)/\eta}}\right\}.
\label{pio2rho2dist_eq}
\end{equation}
We show sample distributions of $P(\rho_0)$ and 
$P({\textstyle{{\pi}\over{2}}}\rho_2)$ in
Figure\ \ref{rhodist_fig}.
As this figure demonstrates,
the distribution of correlation departs from an exponential distribution
at large amplitudes, when the signal-to-noise ratio is
large in a single sample. 

\subsection{Note on Size Estimate From Modulation Index}\label{modind_est_err}

The modulation index, $m=\sqrt{\langle S^2 \rangle/\langle S \rangle^2 - 1}$,
parametrizes source size $\sigma$,
as Eq.\ \ref{modind_size} shows.
One can estimate the modulation index directly using
this relation,
by replacing the ensemble averages with 
finite averages over a set of observations.
Such a procedure replaces  $\langle S \rangle$ 
with ${{1}\over{N}}\sum_{\imath=1}^N S_{\imath}$,
and $ \langle S^2 \rangle$ with 
${{1}\over{N}}\sum_{\imath=1}^N S_{\imath}^2$,
and yields an estimate $\hat m$
for the modulation index.
Unfortunately,
these finite sums are weighted by flux density,
so that they are most sensitive to the scintles
with the greatest flux density.
For a nearly exponential distribution of flux
density, as expected for a small source
in strong scattering, these strong scintillations
are rare. Therefore, the sums
are slow to converge.
The measurement is heavily
subject to the shot noise
of scintillation (\cite{cor98}).
Moreover,
source size affects the weakest parts
of the distribution, to which these sums are least sensitive;
whereas most of the systematic effects discussed
above are greatest for the strongest parts of the distribution.

To exclude the possibility that
the estimate of size is biased by effects 
other than source structure,
such as those discussed in \S\S \ref{variations_gain_flux_sec}
through \ref{average_modind},
a careful study of the full distribution
is the most accurate approach.
The modulation index expresses
the distribution as a single value.
Weakest scintles are also the most common,
so the important parts of the distribution
can be measured with greater accuracy
than can the modulation index.
Although the shot noise of individual scintles
can limit the accuracy of a measurement
based on a fit to the distribution,
the accuracy is much better than provided
by estimates of modulation index.

\subsection{Summary: Expected Distribution of Flux Density}

A pointlike scintillating source produces
an exponential distribution of flux density,
so that one expects an exponential distribution
of visibility amplitude for an interferometer, if the
baseline is short compared to the spatial scale of
the diffraction pattern.
Averaging in frequency
or in time can
reduce the modulation due to scintillation.
We show that 
variations in system gain or intrinsic flux density
of the source affect this distribution,
particularly if the variations are of order 100\%.
System noise and self-noise also change the distribution.
System noise affects the form of the distribution function
at small amplitudes, and by a change in
the normalization at large amplitudes. 
Self-noise 
affects the distribution at all amplitudes when 
the number of samples $N_q$ is small.
Normalization of the correlation function,
and 1-bit sampling, affect the correlation at large amplitudes,
particularly when the source is strong relative to system noise.

For a source of small, but finite size,
the distribution of flux density is the weighted sum of 3 exponentials.
By ``small'' we mean here that the source
is small compared with the resolution
of the scattering disk seen as a lens.
Scales of two of these exponentials depend on the size of the source,
and are thus useful for size determination, the subject of this paper.
These two exponentials 
affect the distribution 
at small flux density.
Because effects of system noise and source size are both
important at small correlation amplitude,
both effects must be included in data reduction.

Considerations of finite size,
noise, self-noise, normalization, and quantization
can be important for single-dish observations of scintillating pulsars as
well. The effects of these factors depend on the 
details of the detection and sampling schemes used at the antenna.
Jenet \& Anderson (1998)\markcite{jen98} describe some of the
important effects, for digitization and autocorrelation.

\section{Observations and Data Reduction}\label{obs_datared_section}

\subsection{Observations and Correlation}\label{obs_corr_sec}

We observed the Vela pulsar and comparison quasars on 1992 Oct 31 to
Nov 1 using radiotelescopes at Tidbinbilla (70 m diameter), Parkes (64
m), and Hobart (25 m) in Australia; Hartebeesthoek (25 m) in South
Africa; and the 7 antennas of the Very Long Baseline Array of the US
National Radio Astronomy 
Observatory\footnote{The National Radio  Astronomy Observatory is
operated by  Associated Universities Inc., under a  cooperative agreement with
the National Science Foundation.}
that could usefully observe the
Vela pulsar. 
Each antenna observed right-circular polarized radiation
in 14$\times$ 2-MHz bands
between 
2.273 and 2.801~GHz.
These bands are sometimes referred to as ``IF bands''
or ``video converters'';
here we call them frequency bands or simply bands.
The data were digitized to 1 bit 
(that is, we measured only the sign of the electric field),
and recorded with the Mark III recording system.

We correlated the data with the Mark IIIB correlator at
Haystack Observatory,
with time resolution of 5~s and 
with 160 time lags.
Fourier transform of the correlation functions
to the frequency domain yields
cross-power spectra with 80 channels, with frequency resolution
of 25~kHz.
Because of correlator limitations, we correlated only
6 of the 14 recorded bands for most scans.
Correlations are tabulated
as normalized fractional correlation
with one-bit sampling, $\rho_2$,
as given by Eqs.\ \ref{rho_0_TA_TS} and\ \ref{van_vleck_relation}.

For each baseline,
we correlated the data in 3 gates across the pulse.
Gate 1 covers
approximately the 13 milliperiods
up to the peak of the pulse,
Gate 2 covers the next 13 milliperiods, and 
Gate 3 covers the next 25 milliperiods.
The 3 gates have average intensity ratios of about 
0.7\,:\,1\,:\,0.5,
as discussed further in \S \ref{obsintdist_sec} below.
The processor set the gates so that they were delayed according
to the pulsar's measured dispersion
(\cite{tay93}, \cite{lyn98})
at the center frequency of each of the 14 bands.
Thus, dispersion smears the gates in time by no more
than the dispersion across 1~MHz,
or $<50~\mu{\rm s}\approx 0.6$~milliperiods.
To the extent that structure of the pulsar's emission
region changes with pulse phase, we can regard
the gates as sampling diffraction patterns from sources with
different structures,
or from a single source viewed from different angles.

The Haystack correlator
is a lag or ``XF'' correlator,
so spectral data are affected
by the fractional bitshift effect (\cite{tms86}).
After correlation,
we removed the average phase slope
introduced by this
effect
for each time sample.
Because the pulsar period is short compared to the fractional
bit-shift rate, we did not encounter aliasing
between the fractional bit-shift and the pulse period (\cite{brith97}).

For each baseline, we analyzed all 3 gates 
with identical phase models, with identical correlator parameters
and the same phase model removed from each gate.
This phase model consists of a phase offset and slopes in
time and frequency,
fit to the data in the 3 gates,
for all correlated tracks over each 13-minute scan.
This process is often called ``fringing'' the data.
Because we study the distribution of 
amplitude of correlation in this paper,
the phase model influences the results only
indirectly, through the effects of averaging.

After fringing,
we averaged the data in time to increase signal-to-noise ratio.
We boxcar-averaged the data by 2 samples, to a resolution of
10~s in time.
This resolution in frequency and time is finer than the
diffraction pattern of the pulsar,
but coarse enough to yield good signal-to-noise ratio.

In this paper we focus on observations 
on the short Tidbinbilla-Parkes baseline, about $200$~km long.
The Tidbinbilla and Parkes 
antennas are both large and have sensitive receivers.
They can be regarded as nearly identical;
any difference is important only for details of the 
distribution of normalized correlation at high amplitude,
discussed in \S\ref{normcorr_vanvleck_sec} above 
and \S\ref{rhodist_obs_sec} below.
Because the characteristic spatial scale of the diffraction pattern is
$\lambda/\theta \approx (8000~{\rm km~East-West})\times (13000~{\rm km~North-South})$,
this short baseline effectively measures the 
flux density of the scintillation pattern at a single point
in the plane of the observer.

\subsection{Instrumental Gain and Noise}\label{ssec_gain_noise}

\subsubsection{Gain from a Continuum Source}

\begin{figure}[t]
% fig 5: Gain with frequency channel as measured for a continuum source
\plotone{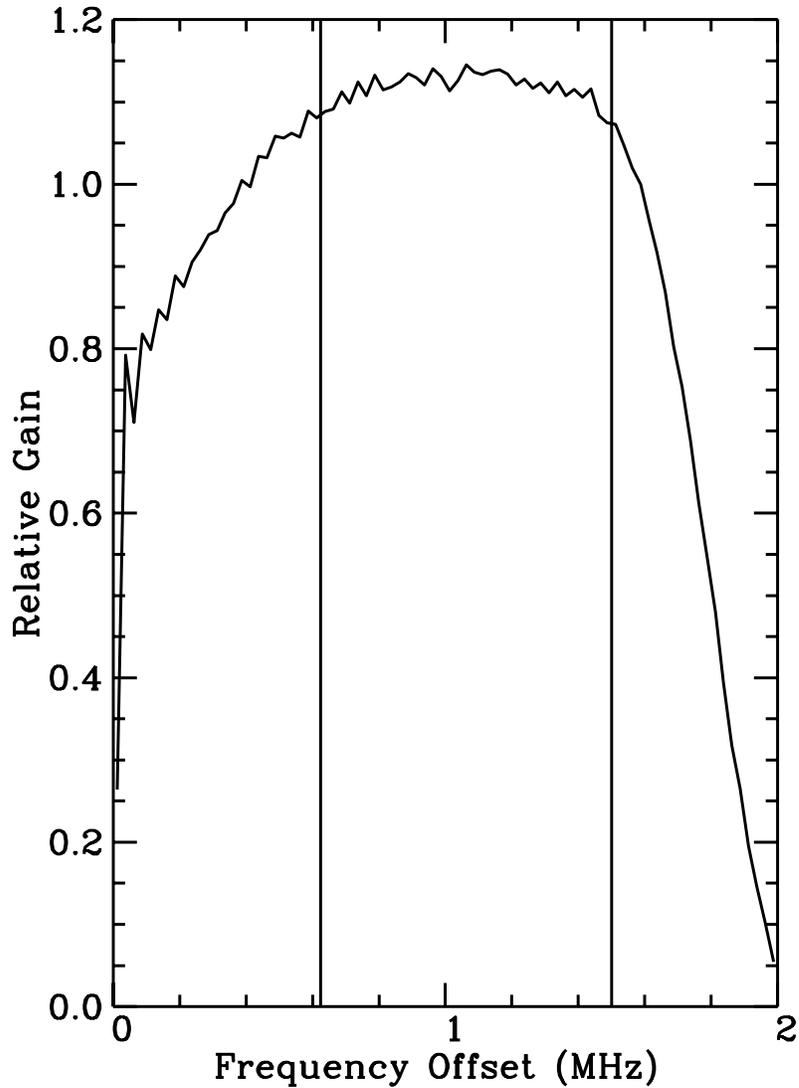}
\figcaption[]{
Gain plotted with frequency
in a typical 2-MHz recorded band.
Data are for the extragalactic continuum source
0826$-$373,
observed from 00:22:50 to 00:35:30 UT
on 1992 Nov 1,
between 2286.99 and 2288.99~MHz
on the Tidbinbilla-Parkes baseline.
The data were
averaged coherently for 10~s in time and 25~kHz in frequency,
and then averaged incoherently in time for the entire 13-minute scan.
To reduce effects of gain variations,
we use only data from
the central portion of the band, between the vertical lines.
\label{passband_fig}}
\end{figure}

From observations of a strong continuum source,
we found that gain varies within each recorded band.
Figure \ref{passband_fig} shows the
amplitude of a strong continuum source plotted with
frequency, as an example.
The source, 0826$-$373,
was observed on 1992 Nov 1 from 0:22:30 UT to 0:35:30 UT.
The figure shows the band
between 2286.99 and 2288.99~MHz.
The rolloff of gain at the high- and low-frequency ends of the band
arises primarily from the filtering required to isolate the recorded
band.
These gain variations, and the differences in gain between
recorded bands,
can be measured from observations of a continuum
source and, in principle, corrected.
However, the noise does not follow the same profile,
so this correction would distort the statistics of
noise.
Because our measurements rely heavily on
accurate knowledge of the noise level,
we delete the outer part of each recorded band
(the lower 25 and the upper 20 of the 80 channels,
as shown in Figure\ \ref{passband_fig}),
and use the central part without gain calibration.
We detect no significant variation of gain with time.

\subsubsection{Noise from a Continuum Source}

We measured the system noise from
the distribution of amplitude for 0826$-$373,
and compared the result with other estimates.
For a strong continuum source, the distribution of correlation
is expected to follow 
a 2-dimensional Gaussian distribution in the complex plane,
offset from the origin by the correlated
flux density of the source,
with standard deviation equal to the noise
level $\sigma$ (\cite{tms86}).
For a strong source, 
the distribution of the amplitude of the correlation
follows a 1-dimensional 
Gaussian distribution with the same standard deviation $\sigma$,
centered on the mean amplitude.

We found the distribution of amplitude about the mean for 
0826$-$373
using the same data shown in Figure\ \ref{passband_fig}.
Each amplitude was measured with bandwidth 25~kHz and time averaging 
of 10~s,
in the central region of the band, as
shown for a single band in Figure \ref{passband_fig}.
We fit a Gaussian distribution to the amplitude found in each
of the 6 correlated frequency bands,
with parameters of the standard deviation,
normalization, and mean amplitude.
Table\ \ref{table1_cont_hist_6_1f2t_002231}
summarizes results of these fits.
The noise, given by standard deviation, should be identical in each band.
The fits show a variation of up to 2.4\% in the noise,
or up to 2.2 times the standard errors.
Again, we do not regard these differences 
as significant.
The mean amplitudes show variations of up to 3.0\%,
with high significance,
reflecting variations in instrumental gain in
the different bands.
The differences in the noise 
do not correlate with the differences in gain.
We could correct for the gain variations,
but prefer to preserve identical conditions
in all bands.

\begin{figure}[t]
% fig 6: histogram of noise after removing amplitude differences
\plotone{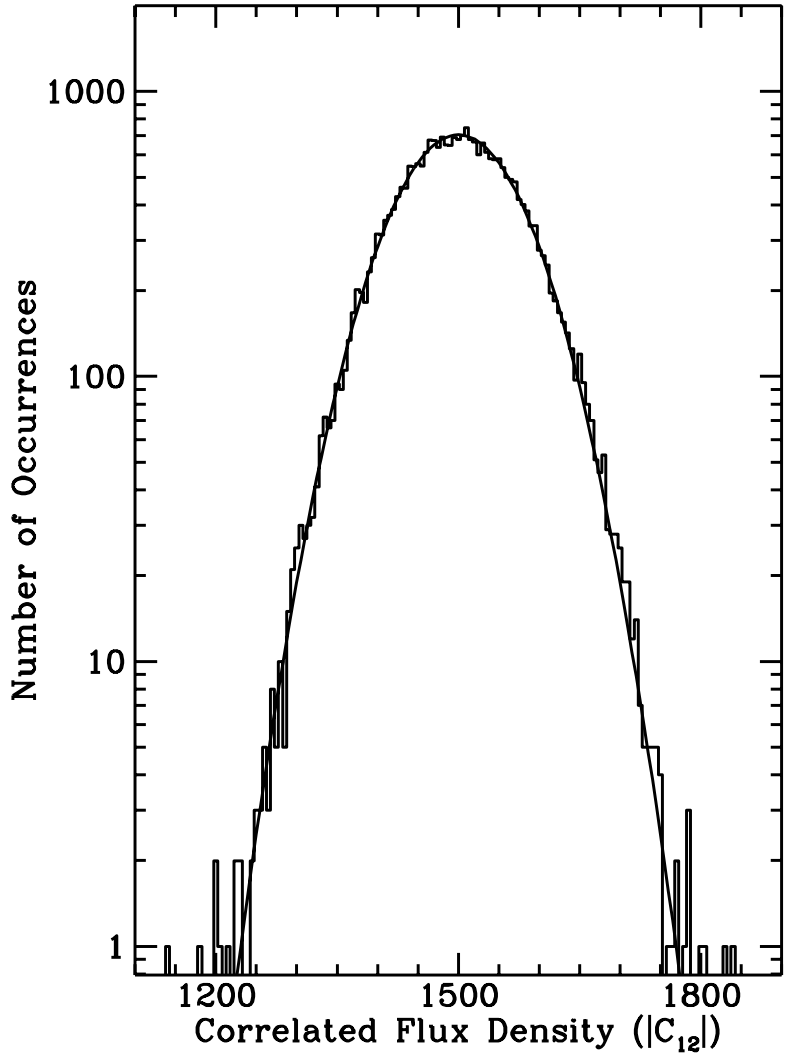}
\figcaption[]{
Histogram of noise.
Data are the same as in Fig.\ \ref{passband_fig}.
To remove the effects of gain variations on the relatively
high amplitude of the source, we subtracted 
the average amplitude given in 
Table\ \ref{table1_cont_hist_6_1f2t_002231}
from the data in each band, 
and then added the overall average amplitude of 1500.
Thus, the histogram reflects
noise rather than gain differences among the bands.
The solid curve shows the best-fitting Gaussian distribution,
with parameters given in
Table\ \ref{table1_cont_hist_6_1f2t_002231}.
\label{cont_hist_off_fig}}
\end{figure}

Figure\ \ref{cont_hist_off_fig}
shows the distribution of correlated flux density for 
all the recorded bands.
So that the histogram reflects noise,
rather than the variations in gain between bands,
we removed the mean amplitude for each band, 
as given in 
Table\ \ref{table1_cont_hist_6_1f2t_002231},
and added the overall mean of 1500.
The distribution of noise is nearly Gaussian,
as expected.
A fit to this composite distribution 
yields noise level $\sigma=74.3\pm 1.0$.
This value is quite consistent with that
found from other scans,
and with the values for individual recorded bands, as 
Table\ \ref{table1_cont_hist_6_1f2t_002231} 
shows.

We can also compare this noise level with that found in
observations that failed to detect any fringes,
which should produce purely noise;
and in observations in a gate off the pulsar pulse.
In either case the distribution function in the complex plane is
a circular Gaussian, centered at the origin (\cite{tms86});
the resulting distribution of amplitude 
follows a Rayleigh distribution.
In earlier work, we described the second approach (\cite{gwi97}),
applied to observations made at the same epoch,
but for a different scan on the pulsar,
correlated on a different date and with a slightly different
correlator configuration.
Expressed in the same units, this procedure yielded a noise level of 
$\sigma = 74.0 \pm 0.5$. 
This differs by less than the standard error
from the value we measure here.
We adopt $\sigma = 74.3$.

\subsubsection{Noise for Pulsar Observations}\label{noise_magnet}

Because the pulsar gate reduces the duty cycle,
we must increase the noise level to reflect the shorter
effective integration time,
for comparisons with gated pulsar observations.
Gates 1 and 2 have duty cycles of 0.013,
and Gate 3 has a duty cycle of 0.021;
the noise increases by a consequent factor of 8.77 or 6.32,
respectively.

Quantization during analog-to-digital conversion affects the noise
level. 
Qualitatively, the presence of strong signals in the recorded
band reduces the noise level in frequency channels where the
signal is weak or absent (\cite{gwi99}).  The ``dithering''
employed to enhance dynamic range in some commercial analog-to-digital
conversions is a closely related phenomenon (\cite{bar93}).  
Quantitatively, the
noise is not distributed evenly over the band, but is 
distributed according to the
product of autocorrelation spectra at the 2 antennas
(which can be inferred from the cross-power spectrum)
plus ``white'' digitization noise.
However, the summed, squared noise (in the absence of self-noise)
remains the same as for zero correlation, or a source without
spectral variation.
We compute the expected noise level for weak signals for each 
time sample of the scintillation spectrum of the pulsar, and after excising outliers, 
use the mean
value for the expected noise level.  The resulting values,
used in our analysis, are given in Table\ \ref{table2_convfunc_fits_3best.out}.
Relatively more conservative or more liberal cuts had little
effect on the results of the fits for size, described below.

\section{Distribution of Intensity for the Vela Pulsar}

\subsection{Observed Intensity Distribution and Fits}\label{obsintdist_sec}

We used the distributions 
of amplitude observed for the pulsar, in the 3
gates across the pulse, in combination with the measured system noise
and the models discussed in \S\ref{theory_section}, to find the size
of the pulsar in the 3 gates. 
We used the ratio $\theta_{\xi}/\theta_{\eta}=1.65$
(\cite{gwi97})
to set the ratio of the scales 
$S_{1\xi}=k M \theta_{\xi}\sigma_s$
and 
$S_{1\xi}=k M \theta_{\eta}\sigma_s$
for the fits. 
Just as for the continuum source 0826$-$373,
the data were averaged for 10~s in time
to yield high signal-to-noise
ratio while remaining within the speckle limit, as discussed in
\S\ref{iss_intro} above. 
Extrapolation from previous, published measurements
yields a decorrelation bandwidth for the pulsar of $\Delta\nu=60$~kHz 
at our observing frequency of 2.3~GHz, and a scintillation timescale of
$t_{ISS}=15$~s
(\cite{bac74}; \cite{rob82}; \cite{cor85}).
From our observations, as discussed in \S\ \ref{avg_time_freq_sec} below,
we infer a decorrelation bandwidth of $\Delta\nu=66$~kHz and a 
timescale of $t_{ISS}=26$~s.
% The finite size of the pulsar may increase the decorrelation bandwidth
% slightly,
% compared to what would be measured for a point source 
% (\cite{IO98}).

\begin{figure}[t]
% fig 7: 
\plotone{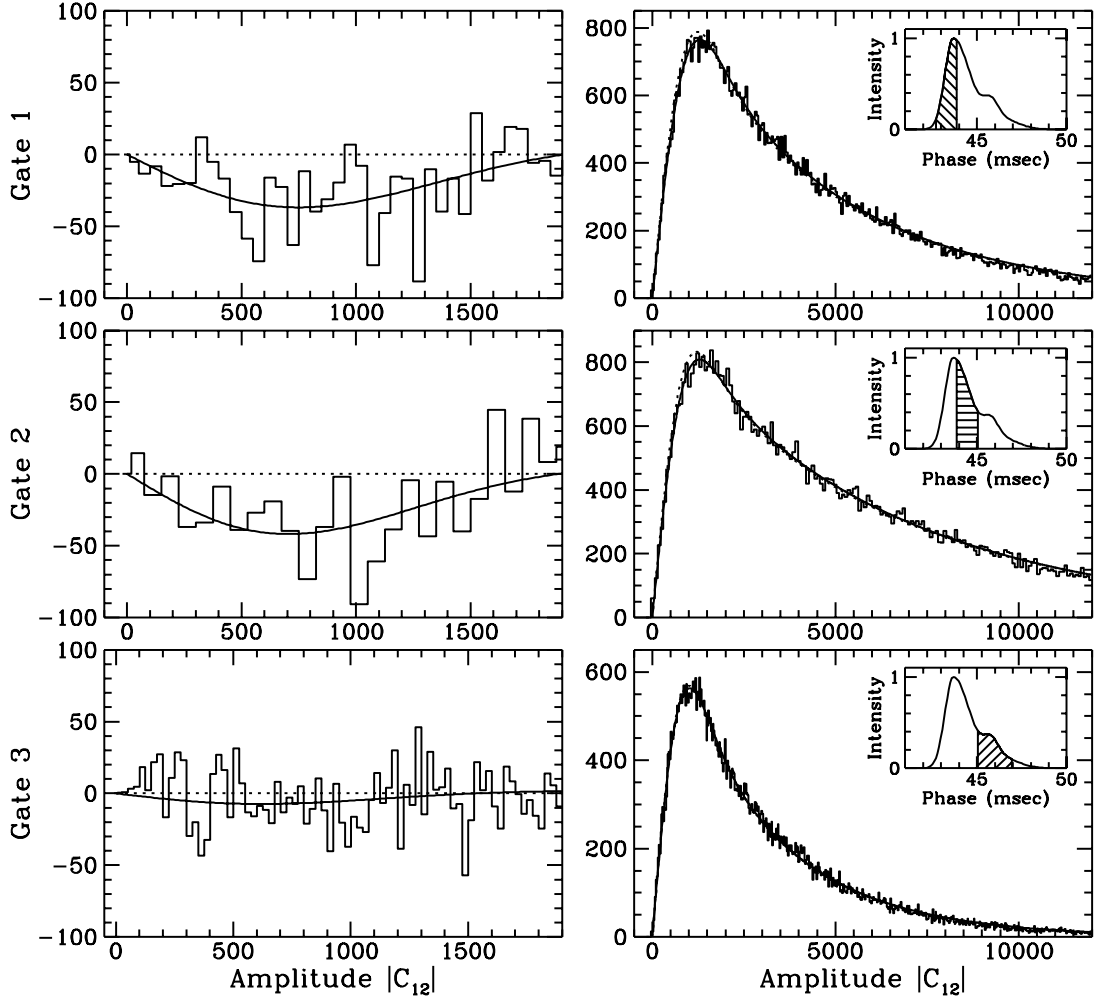}
% gates123.1f1t.binvar.cvf_10.bcut_01.mac
\figcaption[]{
Histograms of observed amplitude in 3 gates across the pulsar pulse.
For each gate, the right panel shows the 
histogram of amplitudes 
with the best-fitting model (solid curve)
and the model with zero size (dotted curve).
Inset 
shows the pulse profile with the gate.
The left panel shows residuals from the model with zero size,
for the data (histogram) and for the best-fitting model (solid curve).
The decrease in pulse intensity, as the gates progress across the pulse,
appears as a decrease in the scale of the exponentials,
from the top panel (Gate 1) to the bottom panel (Gate 3).
\label{3best.gates123.1f2t.binvar_fig}}
\end{figure}

Figure \ref{3best.gates123.1f2t.binvar_fig}
shows the observed distribution of correlation
in the 3 gates.
To each histogram,
we fit a model for the size of the pulsar,
of the form
given by Eq.\ \ref{3Cdist}.
We used the Levenburg-Marquardt algorithm
for the nonlinear fit (see, for example, \cite{pre89}).
We fit for 3 parameters,
corresponding to the overall normalization,
the exponential constant $S_0$,
and the size of the source scaled by scattering parameters, 
$(k M \theta_{\xi} \sigma_s)^2$.
We searched broad ranges of 
initial values for the parameters, including
the range from 0 to 0.3 for the scaled size of the source.
Table\ \ref{table2_convfunc_fits_3best.out}
summarizes the results of the fits.
We found that the best-fitting model
has nonzero size for the pulsar,
with size that decreases across the pulse.
The results for the normalization parameter were 
approximately consistent
with the number of values in the histogram,
as expected for a good fit.

\subsection{Corrections for Averaging in Frequency and Time}\label{avg_time_freq_sec}

\begin{figure}[t]
% fig 8: 
\plotone{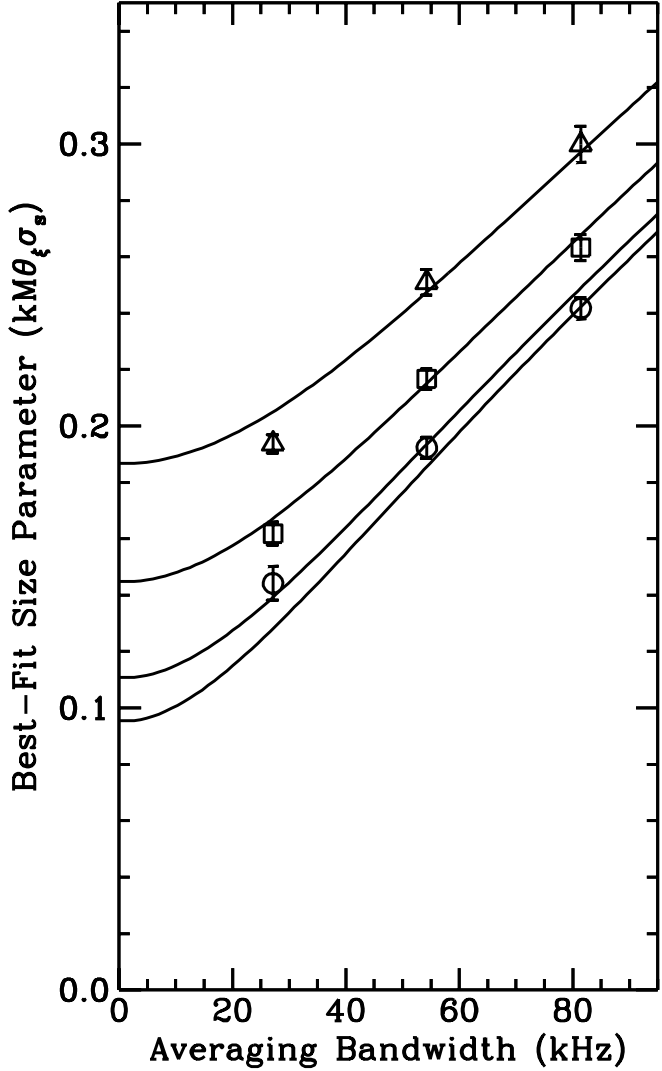}
\figcaption[]{
Fitted size parameter, plotted as a function of 
averaging in time and frequency,
in Gate 2.
The size is normalized to the resolution of the scattering
disk acting as a lens, expressed as $(k M \theta_{\xi} \sigma_s)$.
The ordinate is the averaging bandwidth:
25~kHz as produced by the correlator,
and boxcar averages to 50 and 75~kHz.
Circles show fits for data averaged by 5 sec in time,
squares by 10 sec, and triangles by 15~sec.
The 4 curves show the 
predictions of the best-fitting model 
with parameters of decorrelation bandwidth $\Delta\nu=66$~kHz
and scintillation timescale $t_{ISS}=26$~sec.
The different curves show predictions for averaging
in time
by 0, 5, 10, or 15 sec.
Section\ \ref{average_modind} describes the model.
\label{peakfit_fig}}
\end{figure}

We can find the effects of averaging in time
and frequency by theoretical calculations, 
and by empirically comparing results for data with different
degrees of averaging.
To find results of averaging,
we boxcar-averaged the interferometric correlation
over 1, 2, or 3 of the 25-kHz channels output by the correlator 
in frequency,
and over 1, 2, or 3 of the 5-sec samples in time.
We then fit a model as described in
\S \ref{obsintdist_sec},
with a free parameter for source size,
and with the strength of noise
as found in \S \ref{ssec_gain_noise},
scaled appropriately by the inverse square root of
the averaging time and frequency.
Figure\ \ref{peakfit_fig} shows the results
for Gate 2, the gate in which the
pulsar was strongest.
Results for the other gates are
similar, although the
extrapolated size of the pulsar 
with zero averaging are different.
Averaging in time or
frequency increases the 
best-fitting size,
as expected.

The curves in Figure\ \ref{peakfit_fig}
show the best-fitting forms for the 
expected variation of size ${\bar\sigma_s}$ 
as a function of averaging
bandwidth and time interval,
as given by Eqs.\ \ref{modind_size}, \ref{freq_avg_s2}, 
and \ref{time_avg_s2},
fitted to the data points.
Parameters of the fits are the decorrelation
bandwidth $\Delta\nu$ and scintillation timescale $t_{ISS}$,
and the sizes of the source in the 3 gates
in the absence of averaging.
The best-fitting decorrelation
bandwidth is $\Delta\nu=66\pm 1$~kHz,
and the best-fitting scintillation timescale is
$t_{ISS}=26\pm 1$~sec.
These are comparable to those 
extrapolated from previous measurements,
reported in the literature
(see \S \ref{obsintdist_sec}).
We can remove effects of time and frequency
averaging to find the true sizes for the source
in the 3 gates, without averaging in time or
frequency, inferred from these fits.
Table\ \ref{table2_convfunc_fits_3best.out} gives the results.

Figure\ \ref{3best.gates123.1f2t.binvar_fig}
shows both the best-fitting size,
and the distribution predicted for 
a source of zero size,
averaged in frequency and time
by the same amount. In gates 1 and 2
a model with size greater than zero
fits remarkably better.
This indicates the significance of the fitted size
after correction for time averaging:
about 10~standard deviations in gate 1,
and about 6~standard deviations in gate 2.
In gate 3, a model with size zero fits about as well
as the best-fit, with $k M\theta_{\xi}\sigma_s=0.020$.

\subsection{Gain and Flux Density Variations}\label{gain_flux_variations_sec}

We measure variations of as much as 3.5\% in gain across
the bands, as Figure\ \ref{passband_fig} shows;
and we find variations of gain between bands of as much
as 3.0\%, as shown in Table\ \ref{table1_cont_hist_6_1f2t_002231}.
These are the maximum departures; the distribution is actually
much more concentrated than these figures indicate.
The discussion in \S\ref{variations_gain_flux_sec} shows
that this level of gain variations 
affects the analysis by a completely negligible amount.

Individual pulses from Vela show significant variability
(\cite{kri83}).
These intensity variations are almost completely random in time.
A measure of this variability is the 
the intrinsic, single-pulse modulation index,
$m_{i1}=\langle S_1^2\rangle-\langle S_1\rangle^2/\langle S_1\rangle^2$,
where $S_1$ is the flux of the pulsar in a single pulse.
The index $m_{i1}$ is 0.4 for the average intensity of the pulse.
It varies from 0.3 to 1.3 in individual 
bins of $1^{\circ}$ in pulse longitude,
with the largest variations only at the beginning and end of the pulse,
where the average intensity is low (\cite{kri83}).
Our time averaging over 10~s averages together about 112 pulses.
Our gates cover $4.7^{\circ}$ or $9.0^{\circ}$ of pulse longitude.
The modulation index for an average of $N$ pulses,
assuming that the measurements are uncorrelated, is 
$m_{iN}=(m_{i1})/\sqrt{N}$.
We thus expect the modulation index of our 112-pulse average
to be no more than 0.15, at the very most;
and we expect the overall average of $m_{iN}$ to be closer to 
$0.4/\sqrt{112}=0.04$.

We adopt 15\% as a conservative upper bound for the net effect
of gain variations and intrinsic variability.
Figure\ \ref{superposed_dist_fig} 
indicates that, even if the distribution of intensity
follows a flat distribution with this width,
effects of gain variations and intrinsic variability 
might be barely detectable, over 3 orders of magnitude in
the probability. They will not significantly affect
the measurements of pulsar size.

\subsection{Self Noise}

For our observations, 
the product of bandwidth and integration
time (including the duty cycle of
the pulsar gate) is $N_q=3250$ for Gates 1 and 2, and $N_q=6250$
for Gate 3. 
The average amplitude of the source, $S_0$, is between
12.1 and 4.9 times the noise level, $T_S/(\Gamma \sqrt{N_q})$.  In these
cases, as Figure\ \ref{selfnoisedist_fig} indicates, self-noise has a
negligible effect on the distribution, to our accuracy. In other
words, the distributions are indistinguishable from those for
$N_q\rightarrow \infty$. Self-noise has
negligible impact on the distribution function.

\subsection{Normalized Correlation}\label{rhodist_obs_sec}

\begin{figure}[t]
% fig 9: 
\plotone{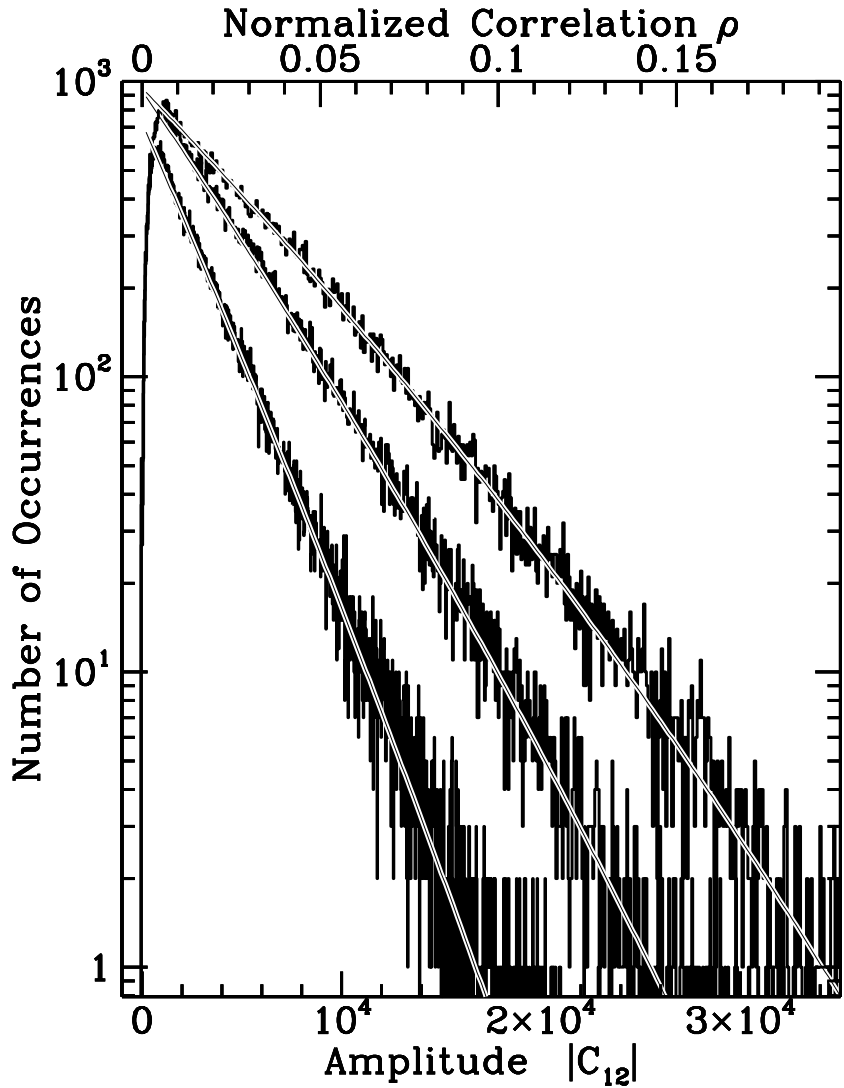}
\figcaption[]{Histogram of distribution of amplitude
in 3 gates,
with the best-fitting distribution
for the distribution of normalized correlation,
as given by Eq.\ \ref{rhodist_eq} and shown in Figure\ \ref{rhodist_fig}.
Table\ \ref{table3_rhofit_table}
gives parameters of the fit.
The distributions are concave downward because of
correlator normalization, as discussed in \S\ \ref{normcorr_vanvleck_sec}.
\label{rhofit_fig}}
\end{figure}

Normalization of the correlation detectably affects the
distribution of amplitude for the Vela pulsar.
Figure\ \ref{rhofit_fig} shows the distributions of amplitude
in the 3 gates, 
with a logarithmic vertical axis.
A purely exponential distribution would follow a straight
line on this plot,
but the observed distribution
falls below this line at both small and large amplitudes.
The figure shows the best-fitting models including effects of normalization
of the correlation (Eq.\ \ref{rhodist_eq})
for each gate, which accurately describe the distributions
at large amplitude.
Table\ \ref{table3_rhofit_table}
summarizes results of the fits
of the distribution expected for a normalized correlation function,
given by Eq.\ \ref{rhodist_eq},
to the data in Fig.\ \ref{rhofit_fig}.
To improve the sensitivity of the fit
to the high end of the distribution,
we fit to the logarithm of the histogram.
The fitted parameters agree well with expectations.
% The Haystack correlator expresses correlations in
% units of $10^{-4}$,
% and an additional factor of $1/\sqrt{160}$ arises in our Fourier transform
% from lag correlation function to cross-power spectrum.
% Therefore the factor for conversion from
% correlation amplitude $|{C}_{12}|$ to
% normalized correlation ${\rho_0}$
% is expected to be about $2\pi/8\times 10^5$.
The normalization is greater than the number of 
data in the histogram, probably because of the reduction 
in the distribution at small amplitudes
by effects of noise and finite source size.

Models including effects of finite size and noise 
(Eq.\ \ref{3Cdist})
accurately describe the distributions at low amplitudes,
where these effects become important,
as Figure\ \ref{3best.gates123.1f2t.binvar_fig} shows.
If we compare the parameters of 
the 2 distributions at the intermediate amplitudes
where both take a nearly exponential form,
using the approximations
in Eqs.\ \ref{erfC_approx} and \ref{rhodist_eq_approx},
unsurprisingly we find that they agree very well,
as Figure\ \ref{rhofit_fig} would suggest.

\section{Discussion}\label{discussion_section}

\subsection{Size of the Pulsar's Emission Region}

Our analysis indicates a decreasing size across the pulse.
Table\ \ref{table4_size_summary} 
summarizes our results,
expressed both in terms of the size parameter,
and as the FWHM of the best-fitting Gaussian distribution,
in km.
To calculate the size in km 
from the results of the fit requires the additional 
parameters $k$, $M$, and $\theta_{\xi}$, $\theta_{\eta}$.
The angular broadening of the pulsar is 
$\theta_{\xi}\times\theta_{\eta}=
(3.3\pm 0.2~{\rm mas})\times (2.0\pm 0.1~{\rm mas})$,
with the major axis at a position angle of $92^{\circ}\pm 10^{\circ}$
(\cite{gwi97}).

We use comparison of our measured decorrelation bandwidth
from \S\ \ref{avg_time_freq_sec},
$\Delta\nu=66\pm 1$~Hz, 
with the angular broadening 
and the pulsar distance of $500\pm 100$~pc (\cite{tay93})
to obtain the characteristic
distance of scattering material
(\cite{des92}, \cite{gwi93}).
We find that the fractional distance of the scattering
material from the Earth to the pulsar is 
$D/(D+R)=0.60\pm 0.05$.
Here recall that $D$ is the distance
from observer to scatterer, and $R$ is the distance
from scatterer to pulsar.
We thus obtain the magnification $M=1.5\pm 0.3$.

We then use the fitted values for the size parameter
$(k M \theta_{\xi} \sigma_s)$ to find the size in km in each gate,
given in 
Table\ \ref{table4_size_summary}.
Note that 
the uncertainty in the magnification factor $M$
dominates the quoted uncertainties of the size.
The uncertainty in the magnification factor stems, in turn, 
primarily from uncertainty in the distance to the pulsar,

The fitted size is in reasonable agreement with
our earlier results for gate 1 (\cite{gwi97}).
The major differences in the analysis are
the accounting for effects of averaging in time and frequency,
for the effects of spectrally-varying signals on noise,
and our use of the measured decorrelation bandwidth 
to determine the magnification factor, in this paper.

\subsection{Size and Emission Mechanism}

Among the 4 classes of processes for pulsar emission
discussed by Melrose (1996)\markcite{mel96}, the
measured size of about 440~km rules out only models 
in which the observed emission comes from close to the polar cap. 
This region has a size of less than a km,
much smaller than the observed size.
If the emission originates at this location,
but is transported to a larger region, 
such models are still viable.

\begin{figure}[t]
% fig 10: 
\plotone{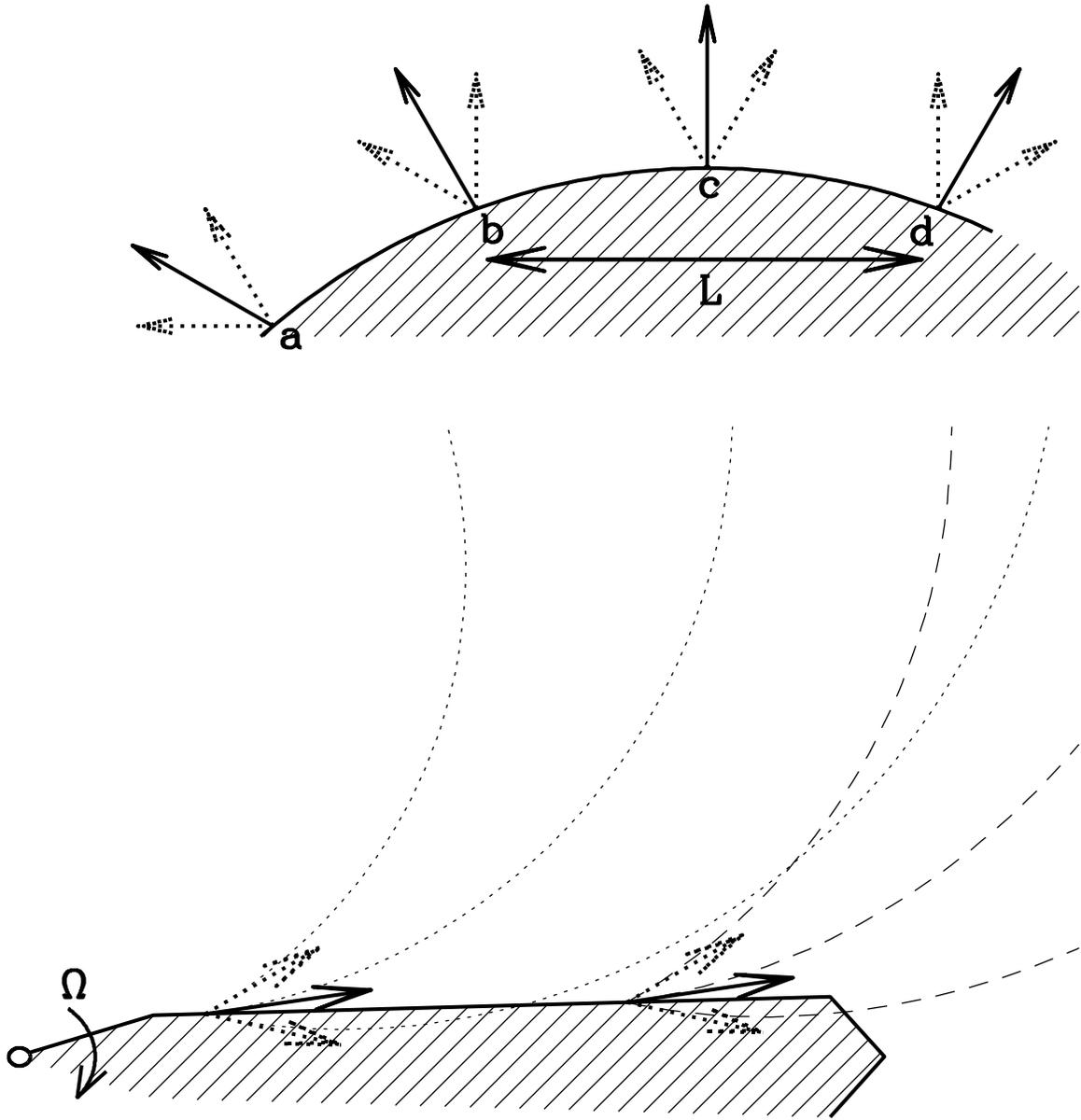}
\figcaption[]{
Upper: Observed size of a source of beamed
radiation.  If the source radiates in a single direction from each
point, an
observer sees only a single point. For example,
if the source radiates only along the solid arrows, an
observer at top sees only 
point ``c''.  If the source emits in a
range of directions at each point, the observer measures a
finite size. For example,
if each point radiates into the angle shown by the dotted
lines, an observer at top sees radiation from
points ``b'' through ``d'' and measures size $L$ for
the emission region. Lower: 
Effects of aberration on measured source size.
The emission region 
rotates
with the pulsar, at left, at angular frequency $\Omega$.
Points on the surface of the emission region radiate
into narrow cones toward the right,
in the rotating frame.
Aberration redirects radiation
so that it travels along the dotted paths from
the lower emitting point, and dashed paths from the upper 
emitting point.
An observer above the source measures a 
size for the emission region
comparable to its height.
\label{weather_retard_fig}}
\end{figure}

Perhaps because pulsar radiation is collimated into a beam, many
pictures of pulsar emission treat the radiation as emerging in only a
single direction from each point of the emitting surface, perhaps
along the local magnetic field.
However, for the emission region to have the 
size we measure, the observer
must receive radiation from points at the emission surface
separated by $\sim 400$~km.
The observer measures the separations between these points
as the size of the emission region.
Figure\ \ref{weather_retard_fig} illustrates this fact.
Of course,
different emitting regions could contribute at different times,
so that the emitting region is a point at any instant,
and the finite size is observed only by averaging over many points.
Our measured size is averaged over 10~s of observations,
or 112 pulses,
and over a range of pulse phase.

Aberration can significantly affect the observed size of the emission
region.  At a fraction of the radius of the light cylinder, the
pulsar's magnetosphere travels at a fraction of lightspeed.  This can
affect the interpretation of the measured size.  In particular, the
measured size can include the height of the emission region.  
Figure\ \ref{weather_retard_fig} shows an example.
Aberration is particularly
important in models where radiation is emitted 
at a range of heights.

Models for emission at a small fraction of the radius of the light
cylinder, $R_L=4250$~km, agree naturally with our observation.
For a simple dipole
field, the open field lines have a cross-section of about 440~km at an
altitude of about 450~km.
However, sources that emit at a range of altitudes
can have measured sizes that include effects of the spread in
altitude,
so that emission could arise at altitudes of 0 to 440~km,
for example.
Models for emission from near or at the light cylinder could
also produce the observed size, although in that case
some physical process
must limit the emission to a region less than 10\% of the diameter
of the light cylinder in size.

The size, and its decrease across the pulse, are 
in approximate agreement with the results
of Krishnamohan \& Downs (1983)\markcite{kri83}.
They concluded that the emission early in
the pulse arose from a larger region than that
late in the pulse, and that the spread in altitude of 
emission was about 400~km.
Although their methodology was quite different,
their picture of the emission region has 
remarkably similar dimensions.

The measured size is large for ``core'' emission,
which is believed to arise near the stellar
surface (\cite{ran90}),
because it statistically reflects
the opening angle of the dipole field lines
at the polar cap.
If Vela's pulse does show core emission,
as some suggest, this leads to an apparent paradox.
The third gate, with size consistent with zero,
might represent the core emission, with the leading
edge of the pulse one side of a cone.
Studies of size of the emission region as a function of frequency,
and imaging of the source,
may resolve this question.
Measurements of size for other pulsars would also be extremely valuable.
Such observations are now in progress.

Arons \& Barnard (1986)\markcite{aro86} pointed out that radiation
traveling nearly along pulsars' magnetic field lines, with
polarization in the plane defined by the pulsar's magnetic field and 
the radiation's wavevector (the ``O-mode''), interacts
with the outflowing electron-positron wind.
This radiation can be ``ducted''
along field lines (\cite{bar86}).  In principle, such
ducting can lead to a large apparent emission region even though the
original source of radiation, perhaps near the pulsar's polar cap, is
quite compact.  
Lyutikov (1998b)\markcite{lyu98b} 
points out that
microphysics can amplify, as well as refract, these waves.

Magnetospheric refraction takes place for only one polarization.
The Vela pulsar is heavily linearly polarized,
so this picture and the size measurements
suggests that the dominant polarization represents the
O-mode.
However, the less-dominant polarization mode is present,
and should show zero size if magnetospheric refraction
sets the size.
Thus, polarimetric size measurements suggest a simple
test of whether magnetospheric refraction is responsible
for the observed size. Such measurements should be little
more difficult than those described here,
and are presently being pursued.

\section{Summary}

We present measurements of the size of the radio emission
region of the Vela pulsar,
from the observed distribution
of correlation amplitude on a short interferometer baseline.
In strong scintillation,
this distribution is exponential if the source is pointlike,
and is the sum of 3 exponentials if the source is small but
extended (\cite{IO98}).
The effects of finite size on
this distribution are concentrated at small amplitudes.
We find that averaging in frequency and time 
has effects similar to those of finite size.
We calculate the distribution 
including effects of Gaussian noise in the observing system,
when an interferometer observes the source.
This effect is likewise greatest at the lowest amplitudes,
with a different functional form.
We calculate the expected effects of variations
in pulsar flux density and in instrumental gain:
these effects tend to make the distribution more 
sharply peaked at zero amplitude and flatter at high
amplitudes, and are small unless the variations
in gain approach 100\%.
We calculate the effects of self-noise, which likewise make the distribution
function higher at low amplitudes and flatter at high amplitudes,
and are important when the number of independent samples,
$N_q$, is small.
We finally consider the effect of normalization of the correlation
function, and find that this makes the distribution function fall
more rapidly than the exponential extrapolated from small
amplitudes, when the correlation approaches 100\%.

We compare model distributions of amplitude with that observed for the
Vela pulsar on the Parkes-Tidbinbilla baseline at $\lambda=13$~cm.  We
measure the noise level, and the variation in gain with frequency
channel, from observations of a strong continuum source.  Models
including a finite size for the pulsar in all 3 gates provide good
fits.  We correct the fitted sizes for the effects of averaging in
frequency and time.  The fitted size parameters are significant at
about 10 times the standard errors, depending on the gate.  Table\
\ref{table2_convfunc_fits_3best.out} summarizes the results.  The size
decreases across the pulse.  We find that effects of gain and pulsar
flux-density variations, self-noise, the Van Vleck correction, and
averaging in time and frequency are expected to be undetectably small
for our observation.  Effects of normalization of the correlation
function are detected, but do
not significantly affect the size estimates.  The linear size of the
pulsar's emission region is about 440~km in the first gate, 340~km in the
second, and less than 200~km in the third.

We discuss size measurements for
sources that, like pulsars, radiate into a collimated beam.  The
measured size is the size of the region on the source visible to the
observer; that size may not be identical from all lines of sight, so
that the measured size may easily change as the source rotates, as we
observe.  Aberration effects may also affect the measured size, so that the
range in altitude of the emission region, as well as its width, may
contribute to the size.

Our measured size is much larger than the sizes of postulated
polar-cap emission regions, so that such models, without additional
physics, are ruled out for this pulsar.  However, magnetospheric
refraction can duct radiation from a compact emission region to
produce a much larger size (\cite{aro86}, \cite{bar86}).  
The observed size is
on the order of postulated emission regions in the lower
open-field-line region,
suggesting that this is a likely location
for pulsar emission. It is much smaller than the outer
open-field-line region or the light cylinder, but emission might arise
in only parts of these regions, or might be beamed toward the observer
from only certain parts.

\acknowledgements

We thank 
B. Rickett for useful discussions
and J. Cordes for sharing an unpublished manuscript.
We thank the U.S. National Science Foundation for financial support (AST-9731584).

\newpage
\figurenum{1}

\newpage
\figurenum{2}

\newpage
\figurenum{3}

\newpage
\figurenum{4} 

\figurenum{5} 
 
\newpage
\figurenum{6} 

\newpage
\figurenum{7}

\newpage
\figurenum{8}

\newpage
\figurenum{9}

\newpage
\figurenum{10}

\newpage

\begin{deluxetable}{lrrr}
\tablenum{1}
\tablewidth{270pt}
\tablecaption{Fits to Noise}
\tablehead{
\colhead{}          & \colhead{}                   &
\colhead{Average}   & \colhead{}                   \\
\colhead{Band}      & \colhead{Noise}              &
\colhead{Amplitude} & \colhead{Normalization\tablenotemark{a}} \\
\colhead{}          & \colhead{$\sigma$}           &
\colhead{$\bar{|C|}$}& \colhead{}                   \\}
\startdata
1 & $ 78.3 \pm 1.2 $ & $ 1459.9 \pm 1.2 $ & $ 4414 \pm 60 $ \\	       
2 & $ 76.8 \pm 1.1 $ & $ 1479.8 \pm 1.0 $ & $ 4387 \pm 52 $ \\	       
3 & $ 75.7 \pm 1.5 $ & $ 1510.4 \pm 1.5 $ & $ 4405 \pm 78 $ \\	       
4 & $ 74.6 \pm 0.8 $ & $ 1524.8 \pm 0.8 $ & $ 4386 \pm 43 $ \\	       
5 & $ 76.6 \pm 1.2 $ & $ 1499.8 \pm 1.1 $ & $ 4384 \pm 58 $ \\	       
6 & $ 73.4 \pm 1.0 $ & $ 1498.5 \pm 0.9 $ & $ 4392 \pm 49 $ \\
All\tablenotemark{b}& $ 74.3 \pm 1.0 $ & $ 1499.6 \pm 1.1 $ & $ 26260 \pm 320 $ \\
\enddata
\tablenotetext{a}{ Actual number of data: 4386 in individual bands, 26316 in all bands.}
\tablenotetext{b}{ Mean of each frequency band subtracted, then 1500 added,
so that distribution reflects noise rather than variations in gain.}
\label{table1_cont_hist_6_1f2t_002231} 
% reference: /home/cgwinn/vela/prod/002231/TP_hists/cont_hist_6_fits_1f2t.tab
\end{deluxetable}

\newpage

\begin{deluxetable}{lcccc}
\tablenum{2}
\tablewidth{490pt}
\tablecaption{Fits to Distributions of Intensity of the Vela Pulsar}
\tablehead{
\colhead{Parameter} &\colhead{}&\colhead{Gate 1} &\colhead{Gate 2} &\colhead{Gate 3} }
\startdata
{\bf Fixed Parameters}&&&&\\
Number of Data                   &                 &$ 74492          $&$ 69305          $&$ 74076          $\\
Bin Width                        &                 &$ 50             $&$ 75             $&$ 25             $\\
Assumed Noise Level\tablenotemark{a}     &                 &$ 558.6          $&$ 507.7          $&$ 474.9          $\\
{\bf Fitted Parameters}&&&&\\
Size Parameter                   & $(kM\theta_{\xi}{\bar \sigma}_s)^2$&$ 0.173 \pm 0.005$&$ 0.163 \pm 0.005$&$ 0.149 \pm 0.005$\\ 
Amplitude                        & $S_0$           &$ 4395  \pm 56   $&$ 6185  \pm 104  $&$ 2826  \pm 23   $\\
Normalization\tablenotemark{b}   &                 &$ 76804 \pm 537  $&$ 72441 \pm 742  $&$ 75631 \pm 340  $\\
Standard Deviation of Residuals\tablenotemark{c}  &&$ 12.29          $&$ 20.49          $&$ 17.99          $\\
&&&&\\
Size Parameter&&&&\\
After Correction for Averaging\tablenotemark{d}&$(kM\theta_{\xi}\sigma_s)^2$&$0.091\pm 0.009$&$0.070\pm 0.012$&$0.020\pm 0.020$\\
\enddata
\tablenotetext{a}{ Corrected for effects of source amplitude as discussed in \S\ \ref{noise_magnet}.}
\tablenotetext{b}{ Normalization larger than the number of data points 
reflects the correction for normalization of the correlation
function (Eq.\ \ref{rhodist_eq}),
not included in the fit. See \S \ref{rhodist_obs_sec} below.}
\tablenotetext{c}{ Differences in the standard deviations of residuals 
among gates
reflect the different widths and populations of bins.}
\tablenotetext{d}{ Assuming decorrelation bandwidth of $\Delta\nu=66$~kHz
and decorrelation time of $t_{ISS}=26$~sec. See \S\ \ref{average_modind} and Fig.\ \ref{peakfit_fig}}
\label{table2_convfunc_fits_3best.out}
\end{deluxetable}

\newpage

\begin{deluxetable}{lcccc}
\tablenum{3}
\tablewidth{470pt}
\tablecaption{Fits for Normalized Correlation}
\tablehead{
\colhead{Parameter}            &                 &\colhead{Gate 1}  &\colhead{Gate 2}  &\colhead{Gate 3}}
\startdata
Conversion factor      &$|C_{12}|/(\rho_0/\eta)$ &$(5.6 \pm 1.4)\times 10^{-6}  $&$(5.6 \pm 1.4)\times 10^{-6}  $&$(5.6 \pm 1.4)\times 10^{-6}  $\\
Normalization\tablenotemark{a} &                 &$77640 \pm 1212   $&$73430 \pm 746   $&$78756\pm 1086$    \\
Scale                          &$\Gamma S_0/T_s$&$0.0233\pm 0.0002 $&$0.0324\pm 0.0004$&$0.0152\pm 0.0001$\\
\enddata
\tablenotetext{a} {Discrepancy from actual number of data, as given in Table\ \ref{table2_convfunc_fits_3best.out}, reflects
effects of noise and pulsar size at low amplitudes.}
% reference /home/condor/cgwinn/vela/prod/TP_hists_22/1f2t/rhodist/rhofit_3.out
\label{table3_rhofit_table}
\end{deluxetable}

\begin{deluxetable}{lcccc}
\tablenum{4}
\tablewidth{470pt}
\tablecaption{
Size of the Vela Pulsar
}
\tablehead{
\colhead{ } & \colhead{ } & \colhead{Gate 1} & \colhead{Gate 2} & \colhead{Gate 3} }
\startdata
Size Parameter&&&&\\
\quad Corrected for Averaging&$(kM\theta_{\xi}\sigma_s)$&$0.091\pm 0.009$&$0.070\pm 0.009$&$0.020\pm 0.020$\\ 
Size (FWHM)\tablenotemark{a} &$\sqrt{8\ln 2}\sigma_s$   &$440\pm 90$~km  &$340\pm 80$~km  &$100\pm 100$~km   \\
\enddata
\tablenotetext{a} { Uncertainties dominated by uncertainty in magnification factor $M$.}
% reference /home/cgwinn/tex/vela/size/size_revised_dnu.nb
\label{table4_size_summary}
\end{deluxetable}
\end{document}